\documentclass[letterpaper]{JHEP3}
\usepackage[applemac]{inputenc}
\usepackage[T1]{fontenc}

\usepackage[dvips]{graphicx}

\usepackage{epsfig}
\usepackage{bbm}

\title{Quasi-hole solutions in finite noncommutative Maxwell-Chern-Simons
theory}

\author{Jules Lambert$^1$ and M. B.  Paranjape$^2$ \\
        Groupe de physique des particules, Département de physique,
Universit\'e de Montr\'eal\\
    C.P.\ 6128, succ.\ centre-ville, Montr\'eal, Qu\'ebec, Canada H3C 3J7\\
    E-mail:
$^1$\email{jules.lambert@umontreal.ca},
$^2$\email{paranj@lps.umontreal.ca}
  }

\preprint{UdeM-GPP-TH-07-158}
\abstract{We study
Maxwell-Chern-Simons theory in 2 noncommutative spatial dimensions
and 1 temporal dimension.  We consider a finite matrix model
obtained by adding a linear boundary field which takes into account
boundary fluctuations.  The pure Chern-Simons has already been
shown to be equivalent to the Laughlin description of the quantum
Hall effect \cite{polychronakos, Hellerman}.   With the addition of
the Maxwell term, we find that there exists a rich spectrum of
excitations including solitons \cite{Isabeau} with nontrivial
``magnetic flux" and quasi-holes with nontrivial ``charges", which
we describe in this article.  The magnetic flux corresponds to
vorticity in the fluid fluctuations while the charges correspond to
sources of fluid fluctuations.  We find that the quasi-hole
solutions exhibit a gap in the spectrum of allowed charge.}

\begin{document}

\pagenumbering{arabic}


\maketitle \tableofcontents

\newpage

\section{Introduction}
In recent years noncommutative geometry has been an effervescent
field of research particularly in its relation to solitons in effective
descriptions of string theory and D-branes \cite{witten-etc}\cite{Madore}.  
However its
most surprising application comes in a description of strongly correlated
quantum magneto-hydrodynamics and various other quantum dynamical fluids
\cite{jackiw-etc}.  The most intriguing application in this context is to
the quantum Hall effect.   Susskind\cite{Susskind} proposed that
non-commutative Chern-Simons theory in  2+1 dimensions would be an
appropriate description of the quantum Hall effect.  The quantum Hall effect
concerns the strongly correlated quantum dynamics of a two dimensional
electron gas in a strong transverse magnetic field.  The noncommutative
space exists in the internal two dimensional space of the Lagrange
coordinate description \cite{Lamb-etc} of the electron fluid.

The continuum, classical description of the small fluctuations of a two
dimensional fluid is easily seen to be a gauge theory of the group of area
preserving diffeomorphisms.  The gauge fields (spatial components)
correspond to fluctuations of the fluid with respect to the ground
state of a quiescent, undisturbed fluid.  The gauge freedom of area
preserving diffeomorphisms, simply corresponds to a relabeling of the
elements of the fluid which are native to the Lagrange description of fluid
dynamics, an evident invariance of the theory.  The corresponding
conservation law is equivalent to the Gauss law.

In the presence of a strong transverse magnetic field and in the low energy
approximation the classical term with the lowest number of derivatives is
exactly the Chern-Simons term.  In this theory, the Gauss law in fact
imposes the vortex free condition on the fluid.  The vortices are frozen out
of the fluid and act as sources, just as ordinary charges in electrodynamics
act as sources outside of the electric and magnetic fields.  Imposing the
Gauss law via the introduction of a temporal gauge fields and an enhanced
gauge invariance (now including time dependent gauge transformations)
results in a fully non-abelian Chern-Simons gauge theory of the group of
area-preserving diffeomorphisms.

Susskind's \cite{Susskind} key observation was that this non-abelian
gauge theory appears to be a truncation to first order of the
simplest noncommutative Chern-Simons gauge theory defined on two
noncommutative spatial and one normal temporal dimension.  Thus
Susskind proposed that the true theory of the quantum Hall effect
corresponds to the full noncommutative Chern-Simons gauge theory.
One motivation given for this hypothesis was to reintroduce the
discreteness that exists at the particulate level of the two
dimensional electron gas, a discreteness which the continuum
approximation erases.  It remained to be seen if the phenomenology
of the quantum Hall effect could be reproduced with this hypothesis.
It was shown in Susskind \cite{Susskind} but also in more detail in
\cite{polychronakos,Hellerman} that indeed the noncommutative gauge
theoretical description of the quantum Hall effect when restricted
to a finite droplet of the fluid through the introduction of a
boundary and boundary degrees of freedom, was in one to one
correspondence to the description afforded by the Laughlin wave
functions \cite{Laughlin}. However, the probability densities
calculated in the noncommutative Chern-Simons model was only equal
to that of the Laughlin wave functions in large distance limit
\cite{karabali}.  This theory however describes the quantum Hall
state via the projection to the lowest Landau level, it cannot hope
to describe any  transition between levels or the transition to the
final state called the Hall insulator \cite{burgess} for very strong
external magnetic field.

In the absence of the transverse magnetic field, the lowest order term in
the effective Lagrangian corresponds to the Maxwell term for the gauge field
of area-preserving gauge transformations.  The coefficient of the
``electric" part does not have to be correlated with the coefficient of the
``magnetic" part combining to give a relativistically invariant action,
however this can be arranged by re-scaling the gauge field or time variable
appropriately.  Again the Gauss law constraint can be obtained by the
incorporation of a temporal component to the gauge field and writing a gauge
invariant expression for the field strength.  The Maxwell term is the next
order term that can be added to the pure Chern-Simons gauge theory.  It
renders the theory more interesting, the Gauss law constraint does not expel
vortices from the theory but imposes a more dynamical constraint.  We have
already studied this theory in previous articles, where we found plane wave
solutions  for the unbounded theory \cite{Garnik} and soliton solutions for
the theory of a finite droplet \cite{Isabeau}.  In this article we further
examine the theory on a finite droplet and show the existence of quasi-hole
states and rotational excitations.  With this rich spectrum of excitations 
we expect that the theory
should be able to describe transitions as a function of the parameters.
Other authors\cite{cappelli} have looked for
the quantum solution of the noncommutative Maxwell-Chern-Simons theory and
found the
correspondance to be to more than one Landau level. However,  in order to
find their
solution they had to assume that a certain deformation energy of the fluid
(defined in the next section) was either zero  or infinity. In this article,
we find classical quasi-hole solutions for the noncommutative
Maxwell-Chern-Simons theory for arbitrary values of the deformation energy.

\section{The model, equations of motion and the Hamiltonian}
Susskind's\cite{Susskind} idea was to describe a two (spatial) dimensional
fluid by a gauge field $A_{j}$ so that
\begin{equation}
x^{i}=y^{i}+\theta\epsilon^{ij}A_{j}\label{coordone gauge field}
\end{equation}
where $x_{i},\,\, i=1,2$ are the Eulerian coordinates of the fluid,
$y_{i},\,\,  i=1,2$
are the Lagrangian (comoving) coordinates of the fluid and $\theta = 1/(2\pi\rho_0)$. Then the
Lagrangian of a charged fluid in an external transverse magnetic field
corresponds to the 2+1 dimensional Maxwell-Chern-Simons theory for small
value of the gauge field. The continuum approximation removes the
discreteness that is manifest in the physical fluid.  Susskind proposed to
bring this discreteness back by suggesting that the noncommutative version
of this theory should describe the full theory.  We studied this theory with
the additional modification of boundary and boundary degrees of
freedom\cite{Isabeau}.  Here we study the same action, however we add a
factor $\kappa$
to the analog of the magnetic field squared term which corresponds to the
potential energy density of spatial deformations of the fluid. The first 
term, the
analog of the electric field squared,  corresponds to the kinetic term,
the Chern-Simons term represents the interaction of the charged fluid with the
external magnetic and the last term represents the boundary degrees of
freedom:
\begin{eqnarray}
\lefteqn{S = \frac{\pi}{g^{2}\theta}\int
dt(Tr\{(-2[D_{0},D][D_{0},D^{\dagger}]-\kappa[D,D^{\dagger}][D,D^{\dagger}]))}\nonumber\\
& & +2\lambda(-[D,D^{\dagger}]+1)D_{0}\}-2\Psi^{\dagger}D_{0}\Psi) .
\end{eqnarray}
Here $D_{0}$ is the time covariante derivative, $D$ and
$D^{\dagger}$ are the holomorphic and anti-holomorphic combinations of
the spatial covariant derivatives respectively and $\Psi$ is a boundary
field.  The boundary field was first added by Polychronakos
\cite{polychronakos} which allowed him to  find solutions of non-commutative
Chern-Simons theory in terms of finite matrices.  These correspond to finite
droplets of the quantum Hall fluid.  Specifically
\begin{equation}
D_{\mu}= \sqrt{\theta}(-i\partial_{\mu} + A_{\mu})
\end{equation}
and
\begin{equation}
D=\frac{D_{1} + iD_{2}}{\sqrt{2}}, D^{\dagger}=\frac{D_{1}
-iD_{2}}{\sqrt{2}}
\end{equation}
and the parameters $\lambda$ and $g^2$ are related to the noncommutativity
parameter $\theta$, the electron charge $e$,  the external magnetic field 
$B$,
the density $\rho_0$ and the electron mass $m$ by
\begin{equation}
\lambda=\frac{eB\theta^{1/2}}{m},\quad g^{2}=\frac{(2\pi)^{2}\rho_{0}}{m} .
\end{equation}
We rescale $D_{0}$ and $\Psi$ and the parameters in the following way, in
order to obtain exactly the action studied in \cite{Isabeau}:
\begin{equation}
D_{0}= \sqrt{\frac{\theta}{\kappa}}(-i\partial_{0} +
A_{0}),\quad \lambda=\frac{eB\theta^{1/2}}{m\sqrt{\kappa}},\quad
g^{2}=\frac{(2\pi)^{2}\rho_{0}}{m\kappa},\quad
\Psi\rightarrow\frac{\Psi}{\sqrt[4]{\kappa}}
\end{equation}
Defining
\begin{equation}
\Xi=\frac{\pi}{g^{2}\theta}
\end{equation}
we obtain the action
\begin{eqnarray}
\lefteqn{S = \Xi\int
dt(Tr\{(-2[D_{0},D][D_{0},D^{\dagger}]-[D,D^{\dagger}][D,D^{\dagger}]))}\nonumber\\
& & +2\lambda(-[D,D^{\dagger}]+1)D_{0}\}-2\Psi^{\dagger}D_{0}\Psi).
\end{eqnarray}
By varying with respect to $\Psi^{\dagger}$, $D_{0}$, and $D^{\dagger}$ we 
get, respectively,  the boundary equation
\begin{equation}
i\dot{\Psi}=A_{0}\Psi\label{equation de psi},
\end{equation}
the Gauss law
\begin{equation}
[D,[D_{0},D^{\dagger}]]+[D^{\dagger},[D_{0},D]]+\lambda([D,D^{\dagger}]-1)+\Psi\Psi^{\dagger}=0,\label{gauss}
\end{equation}
and the Amp\`{e}re law
\begin{equation}
[D_{0},[D_{0},D]]+[D,[D,D^{\dagger}]]=\lambda[D_{0},D].\label{ampere}
\end{equation}
The Hamiltonian is, as in  \cite{Isabeau}, given by
\begin{equation}
H=\Xi\,\,
Tr(-2[D_{0},D^{\dagger}][D_{0},D]+[D,D^{\dagger}][D,D^{\dagger}])\label{hamiltonien_general}.
\end{equation}

\section{Rotational excitations}
Our first solution corresponds to rotational excitations on top of any given
solution.  Our procedure can be applied to the soliton solutions found for
example  in \cite{Isabeau} and to the solutions that we find in this
article. We put
\begin{equation}
D=D'+\frac{R}{\sqrt{2\theta}}e^{i w
t},\,\,D_{0}=D_{0}',\,\,\Psi=\Psi'\label{rotational}
\end{equation}
where the primed variables correspond to any known solution to the equations
of motion with $R$ (proportional to the identity) and $w$ (real) simply
constant.  The Gauss law involves commutators of the $D$ or $D^\dagger$ with
each other or with their commutator with $D_0$.   The direct addition of
complex constants as in the equation (\ref{rotational}) or those that result 
from
the commutators involving $D_0$, simply vanish open taking the further 
commutators
with $D$ or $D^\dagger$, hence the Gauss law is satisfied.   The equation
(\ref{equation de psi}) is also obviously satisfied.  Replacing
(\ref{rotational}) into the Amp\`{e}re law yields a solution if
$w=\frac{\lambda\sqrt{\kappa}}{\sqrt{\theta}}=\frac{eB}{m}$ the familiar 
cyclotron frequency.  We
calculate the corresponding change of energy:
\begin{equation}
\Delta H =-2\Xi Tr\left([D_{0},\frac{R^{*}}{\sqrt{2\theta}} e^{-i w
t}][D_{0},D']+[D_{0},D'^{\dagger}][D_{0},\frac{R}{\sqrt{2\theta}}
e^{i w t}]+[D_{0},\frac{R^{*}}{\sqrt{2\theta}} e^{-i w
t}][D_{0},\frac{R}{\sqrt{2\theta}} e^{i w t}]\right)
\end{equation}
which yields
\begin{equation}
\Delta H =-2\Xi Tr\left(-\lambda \frac{R^{*}}{\sqrt{2\theta}} e^{-i
w t}[D_{0},D']+\lambda \frac{R}{\sqrt{2\theta}} e^{i w
t}[D_{0},D'^{\dagger}]-\lambda^{2}\frac{R^{*}R}{2\theta}\right).
\end{equation}
Since the commutator $[D_{0},D]$ is off-diagonal in our solution
and the solution in \cite{Isabeau} the trace vanishes giving
\begin{equation}
\Delta H =\frac{2\Xi \lambda^{2}|R|^{2}N}{2\theta}.
\end{equation}
Expressing this in terms of physical constants gives
\begin{equation}
\Delta H =\frac{e^{2}B^{2}|R|^{2}N}{2m}
\end{equation}
which is exactly the energy of $N$ rotating electrons with a amplitude $R$
at the frequency $w$ in a magnetic field.

This solution adds a term to the covariant derivative
\begin{equation}
D_{1}=D_{1}'+\frac{|R|}{\sqrt{\theta}}
\cos(wt+\varphi),D_{2}=D_{2}'+\frac{| R|
}{\sqrt{\theta}}\sin(wt+\varphi)
\end{equation}
where $\varphi$ is the phase of $R$. Using the Susskind
correspondence between the fluid coordinates and the gauge field
(\ref{coordone gauge field}), we see that the solution
(\ref{rotational}) corresponds to rotating about the original
solution as
\begin{equation}
y_{1}=y_{1}'-| R|\sin(wt+\varphi),\,\,
y_{2}=y_{2}'+| R| \cos(wt+\varphi)
\end{equation}
where $y_{1}'$ and $y_{2}'$ are for operator coordinates of original
solution. These rotational excitations clearly also exist in the analogous
theory in
the infinite plane treated in \cite{Garnik}.

\section{Quasi-hole solutions}
Quasi-hole solutions appears as a modification of the solutions found in
\cite{Isabeau}.
We will take an ansatz similar to \cite{Isabeau}, hence $D$ is represented
by an $N\times N$ matrix which satisfies certain boundary conditions,
however we will take a periodic $D$ as in \cite{polychronakos}.  The $D$ is
an operator similar in structure to the annihilation operator of an ordinary
Heisenberg algebra, hence it generally relates a state $|n+1\rangle$ to a
state $|n\rangle$.  For the finite matrix representations used here,
periodicity means that the final state $|0\rangle$ is related to the state
$|N-1\rangle$ since $n$ ranges over the $N$ values $0,1,\cdots N-1$.
\begin{equation}
D=\sum_{n=0}^{N-2}\sqrt{G(n)+q}e^{\frac{i
w(n)\sqrt{\kappa}t}{\sqrt{\theta}}}\mid n\rangle\langle n+1\mid+
\sqrt{q}e^{\frac{i \rho \sqrt{\kappa}t}{\sqrt{\theta}}}\mid
N-1\rangle\langle 0\mid\label{anszat D}
\end{equation}
If $q=0$, this ansatz is equivalent to the one in
\cite{Isabeau}.  $G(n)$ is to be determined from the equations of motion, we
solve these eventually, perturbatively and numerically.

Solutions with this ansatz correspond to quasi-hole solutions
because they bound the lowest eigenvalue of the radius away from zero in a
$q$ dependent manner.  From (\ref{coordone gauge field}), the fluctuation of 
the radius is
proportional to the square of the gauge
field
\begin{equation}
A_{1}^{2}+A_{2}^{2}=\left(\frac{D_{1}}{\sqrt{\theta}}+i\partial_{1}\right)^{2}+\left(\frac{D_{2}}{\sqrt{\theta}}+i\partial_{2}\right)^{2}.
\end{equation}
Noncommutative geometry in a finite space is defined by the
commutator of the coordinates $[x_{1},x_{2}]=i\theta(1-N\mid
N-1\rangle\langle N-1\mid)$ which imply the commutator for the
derivate $[\partial_{1},\partial_{2}]=\frac{i}{\theta}(-1+N\mid
N-1\rangle\langle N-1\mid)$. Thus we can put
\begin{equation}
\partial_{1}=\frac{i}{\sqrt{2\theta}}(d+d^{\dagger}),\,\,\partial_{2}=\frac{1}{\sqrt{2\theta}}(d-d^{\dagger})
\end{equation}
where
\begin{equation}
d=\sum_{n=0}^{N-2}\sqrt{n+1}\mid n\rangle\langle n+1\mid .
\end{equation}
Then we obtain
\begin{equation}
R^{2}\propto \{D -d,D^{\dag}-d^{\dagger}\}
\end{equation}
where $\{D^{\dagger},d\}$ is the anti-commutator.
Then the radius becomes
\begin{eqnarray}
\lefteqn{R^{2}\propto
\left(G(0)+2q-1-2\sqrt{G(0)+q}\cos\left(\frac{w(0)\sqrt{\kappa}t}{\sqrt{\theta}}\right)\right)\mid
0\rangle\langle 0\mid}\nonumber\\
& &
+\sum_{n=1}^{N-2}\left(\underline{G(n)}+2q-2n-1-\underline{2\sqrt{G(n)+q}\sqrt{n+1}\cos\left(\frac{w(n)\sqrt{\kappa}t}{\sqrt{\theta}}\right)}\right)\mid
n\rangle\langle n\mid \nonumber\\
& &
+\left(G(N-2)+2q-N+1-2\sqrt{G(N-2)+q}\sqrt{N-1}\cos\left(\frac{w(N-2)\sqrt{\kappa}t}{\sqrt{\theta}}\right)\right)\mid
N-1\rangle\langle N-1\nonumber\\
\mid\label{R2}
\end{eqnarray}
a diagonal expression in the states where
$\underline{A(n)}=A(n)+A(n-1)$. We see that for large $q$ we have
correspondingly large eigenvalues for $R^2$.  The smallest
eigenvalue is not directly equal to $q$, hence our solution
corresponds to the fluid pushed away from the origin.

Returning to the solution of the equations of motion, we will consider the
gauge $A_{0}=0$ but we let $D$ depend on time.   This choice is different
from that taken in \cite{Isabeau}, however in that case,  our choice is
simply gauge equivalent.  With the periodic ansatz of equation (\ref{anszat
D}), this is not the
case.   Indeed, our choice give us an additional degree of freedom which we
can identify as $\rho$.  Thus
$D_{0}=-i\sqrt{\theta}\partial_{0}$, and we can calculate the different
terms in the equations of
motion:
\begin{equation}
[D_{0},D]=\sum_{n=0}^{N-2}w(n)\sqrt{G(n)+q}e^{\frac{i
w(n)\sqrt{\kappa}t}{\sqrt{\theta}}}\mid n\rangle\langle n+1\mid+
\rho\sqrt{q}e^{\frac{i \rho\sqrt{\kappa} t}{\sqrt{\theta}}}\mid
N-1\rangle\langle 0\mid\label{D0Ddag}
\end{equation}
\begin{equation}
[D_{0},[D_{0},D]]=\sum_{n=0}^{N-2}(w(n))^2\sqrt{G(n)+q}e^{\frac{i
w(n)\sqrt{\kappa}t}{\sqrt{\theta}}}\mid n\rangle\langle n+1\mid+
\rho^2\sqrt{q}e^{\frac{i \rho\sqrt{\kappa} t}{\sqrt{\theta}}}\mid
N-1\rangle\langle 0\mid
\end{equation}
The commutator
$[D,D^{\dagger}]$ is given by replacing from equation (\ref{anszat D})
\begin{eqnarray}
\lefteqn{[D,D^{\dagger}]=\sum_{n=0}^{N-2}\sum_{m=0}^{N-2}\sqrt{G(n)+q}\sqrt{G(m)+q}e^{\frac{i
\sqrt{\kappa}(w(n)-w(m))t}{\sqrt{\theta}}} [\mid n\rangle\langle
n+1\mid,\mid m+1\rangle\langle m\mid]}\nonumber\\
& & +q[\mid N-1\rangle\langle 0\mid,\mid 0\rangle\langle N-1\mid]\nonumber\\
& & =\sum_{n=0}^{N-2}(G(n)+q)\mid n\rangle\langle
n\mid-\sum_{n=1}^{N-1}(G(n-1)+q)\mid n\rangle\langle
n\mid+q(\mid N-1\rangle\langle N-1\mid-\mid 0\rangle\langle
0\mid)\nonumber\\
& & = \sum_{n=0}^{N-1}\overline{G(n-1)}\mid n\rangle\langle
n\mid\label{DDdag}
\end{eqnarray}
with the notation $\overline{A(n)}\equiv A(n+1)-A(n)$ and we define
$G(-1)\equiv0$
and $G(N-1)\equiv 0 $.
The commutator $[D^{\dagger},[D_{0},D]]$ can be computed in the same
way as $[D,D^{\dagger}]$.  Defining $w(-1)\equiv\rho$ and
$w(N-1)\equiv\rho$ then
\begin{equation}
[D^{\dagger},[D_{0},D]]=-\sum_{n=0}^{N-1}\overline{(G(n-1)+q)w(n-1)}\mid
n\rangle\langle n\mid .
\end{equation}
As this result is hermitian it is also equal to
$[D,[D_{0},D^{\dagger}]]$. The final term appearing in the  equations of
motion is calculated as
\begin{eqnarray}
\lefteqn{[D,[D,D^{\dagger}]]=\sum_{n=0}^{N-2}\sum_{m=0}^{N-1}\sqrt{G(n)+q}e^{\frac{i
w(n)\sqrt{\kappa}t}{\sqrt{\theta}} }\overline{G(m-1)}[\mid n\rangle\langle
n+1\mid,\mid m\rangle\langle m\mid]}\nonumber\\
& & +\sqrt{q}e^{\frac{i \rho\sqrt{\kappa}
t}{\sqrt{\theta}}}\sum_{m=0}^{N-1}\overline{G(m-1)}[\mid N-1\rangle\langle
0\mid, \mid m\rangle\langle m\mid]\nonumber\\
& & =\sum_{n=0}^{N-2}\sqrt{G(n)+q}e^{\frac{i
w(n)\sqrt{\kappa}t}{\sqrt{\theta}} }\overline{\overline{G(n-1)}}\mid
n\rangle\langle n+1\mid+\sqrt{q}e^{\frac{i \rho\sqrt{\kappa}
t}{\sqrt{\theta}}}\left(\overline{G(-1)}-\overline{G(N-2)}\right)\mid
N-1\rangle\langle
0\mid\nonumber\\
\end{eqnarray}
with the notation
$\overline{\overline{G(n-1)}}\equiv\nabla^{2}G(n)=G(n+1)-2G(n)+G(n-1)$
where $\nabla^{2}G(n)$ is discrete Laplacian (which is facilitated with the
further notational convenience $G(N)\equiv G(0)$ and $\langle N\mid \equiv
\langle 0\mid$).  Then
\begin{equation}
[D,[D,D^{\dagger}]]=\sum_{n=0}^{N-1}\sqrt{G(n)+q}e^{\frac{iw(n)\sqrt{\kappa}t}{\sqrt{\theta}}
}\nabla^{2}G(n)\mid n\rangle\langle n+1\mid .
\end{equation}
The solution for $\Psi$ is simply a general static vector since
$A_{0}=0$ in (\ref{equation de psi}):
\begin{equation}
\Psi=\sum_{n=0}^{N-1}\lambda_{n}\mid n\rangle .
\end{equation}
We see, as in \cite{Isabeau}, that only $\Psi\Psi^{\dagger}$ contributes off
diagonal terms in the Gauss law (\ref{gauss}).  To eliminate such terms we
must take $\Psi=\lambda_{M}\mid M\rangle$.  Contrary to \cite{Isabeau},
different
choices of $M$ are all gauge equivalent (i.e. a permutation) since
all choices of $M$ are equivalent in our periodic ansatz (up to the name of
the
variable).  Taking the trace of the
Gauss law then yields $\lambda_{M}=\sqrt{N\lambda}$. We choose without loss
of generality that $M=N-1$. Then the Gauss law yields (for $n=0, \cdots
N-1$)
\begin{equation}
-2\overline{(G(n-1)+q)w(n-1)}+\lambda\left(\overline{G(n-1)}-1\right)+\lambda
N\delta_{N-1, n}=0\label{gauss en n}.
\end{equation}
We will solve this equation by induction as in \cite{Isabeau}. We
will show that for $n=[0,N-2]$
\begin{equation}
w(n)=-\lambda\frac{n+1-\frac{2q\rho}{\lambda}-G(n)}{2(G(n)+q)}\label{w(n)}.
\end{equation}
This formula  is true for $n=N-2$, as is verified by considering the Gauss
law  (\ref{gauss en n}) for
$n=N-1$. Then assuming the form (\ref{w(n)}) for a general value of $n$ we
can prove that it is valid for $n\rightarrow n-1$.  Thus by the principle of
induction the formula is valid for all $n$.  However we have
to check/impose that the $n=0$ equation of (\ref{gauss en n}) is respected.
This is
indeed the case if we take (\ref{w(n)}).  Taking the Amp\`{e}re law
(\ref{ampere}) and removing an overall factor
$\sqrt{G(n)+q}e^{\frac{i w(n)\sqrt{\kappa}t}{\sqrt{\theta}}}$
we obtain, for $n=[0,N-1]$.
\begin{equation}
(w(n))^{2}+\nabla^{2}G(n)=\lambda w(n)\label{numeric}.
\end{equation}
Here we have $N$ equations, in $N-1$ values of $G(n)$ and also in the two
variables $q$ and $\rho$.  Hence we have $N$ equations and $N+1$ parameters,
thus we have one free parameter.  Generically, there will be a family of
solutions.
The Hamiltonian, from equation \ref{hamiltonien_general}, for this ansatz,
using equations \ref{D0Ddag} and \ref{DDdag} is
\begin{equation}
H=\Xi
(\sum_{n=0}^{N-1}2(w(n))^{2}(G(n)+q)+\left(\overline{G(n-1)})^{2}\right)\label{hamiltonien}.
\end{equation}

\subsection{Perturbative analysis of the quasi-hole solution}

\subsubsection{Large quasi-hole solution}

\label{section large q}

The equation (\ref{numeric}) is non-linear and thus difficult to
solve.  So we will look the solution for $q\gg1$. We define
\begin{equation}
\rho= \rho_{0}+\frac{\rho_{1}}{q}, \,\,G(n)=G_{0}(n) +
\frac{G_{1}(n)}{q}
\end{equation}
Then if we expand $w$ to first order in $1/q$, we obtain:
\begin{equation}
w(n)\simeq\rho_{0}-\frac{1}{2q}(\lambda(n+1-G_{0}(n))-2\rho_{1}+2\rho_{0}G_{0}(n))
\end{equation}
\begin{equation}
(w(n))^{2}\simeq\rho_{0}^{2}-\frac{\rho_{0}}{q}(\lambda(n+1-G_{0}(n))-2\rho_{1}+2\rho_{0}G_{0}(n))\label{wCarre}
\end{equation}

Working to zero order in $1/ q$, equation \ref{numeric} gives  (for
$n=[0,N-2]$)
\begin{equation}
\rho_{0}^{2}+\nabla^{2}G_{0}(n)=\lambda\rho_{0}.
\end{equation}
The discrete Laplacian solved as in the continuous case by
\begin{equation}
G_{0}(n)=\alpha_{0}n^2+\beta_{0}n+\delta_{0},\,\,
\alpha_{0}=\frac{\lambda\rho_{0}-\rho_{0}^{2}}{2}.
\end{equation}
We have to impose the two boundary conditions $G_{0}(-1)=G_{0}(N-1)=0$.
These
give
\begin{equation}
\alpha_{0}-\beta_{0}+\delta_{0}=0,\,\,\alpha_{0}(N-1)^2+\beta_{0}(N-1)+\delta_{0}=0
\end{equation}
which are solved by
\begin{equation}
\beta_{0}=\frac{1}{2}(N-2)(\rho_{0}^2-\lambda\rho_{0}),\,\,\delta_{0}=\frac{1}{2}(N-1)(\rho_{0}^2-\lambda\rho_{0}).
\end{equation}
We still have one final condition left (from the $n=N-1$ of equation
\ref{numeric})  which gives
\begin{equation}
\rho_{0}^2+\alpha_{0}(N-2)^2+\beta_{0}(N-2)+2\delta_{0}=\lambda\rho_{0}
\end{equation}
with solutions
\begin{equation}
\rho_{0}=0 \qquad {\rm or} \qquad \rho_{0}=\lambda .
\end{equation}
Either of these solutions give us $G_{0}(n)=0$.

With the order zero solutions we can continue to  solve the equation
(\ref{numeric}) in first order in $1/q$
\begin{equation}
-\rho_{0}(\lambda(n+1)-2\rho_{1})+\nabla^{2}G_{1}(n)=\frac{-\lambda}{2}(\lambda(n+1)-2\rho_{1}).
\end{equation}
This is solved  by
\begin{equation}
G_{1}(n)=\alpha_{1}n^{3}+\beta_{1}n^{2}+\delta_{1}n+\gamma_{1},\,\,
\alpha_{1}=\frac{\lambda}{12}(2\rho_{0}-\lambda),\,\,\beta_{1}=\frac{1}{4}(2\rho_{0}-\lambda)(\lambda-2\rho_{1}).
\end{equation}
Again we have our boundary conditions $G_{1}(-1)=G_{1}(N-1)=0$
which imply
\begin{equation}
\delta_{1}=\frac{1}{12}(N^{2}\lambda-3\lambda-6N\rho_{1}+12\rho_{1})(\lambda-2\rho_{0}),\gamma_{1}=-\frac{1}{12}(N-1)(\lambda
N-6\rho_{1}+\lambda)(2\rho_{0}-\lambda).
\end{equation}
Finally the condition for $n=N-1$ of (\ref{numeric}) gives
\begin{equation}
\rho_{1}=\frac{1}{4}(N-1)\lambda .
\end{equation}
Thus
\begin{eqnarray}
\delta_{1}&=&-\frac{\lambda}{24}(N^{2}-9N+12)(\lambda-2\rho_{0})\\
\gamma_{1}&=&\frac{\lambda}{24}(N-1)(N-5)(2\rho_{0}-\lambda)\\
\beta_{1}&=&\frac{\lambda}{8}(3-N)(2\rho_{0}-\lambda)
\end{eqnarray}
From the Hamiltonian (\ref{hamiltonien}),  for the $\rho_{0}=0$ solution, it
can be shown with a little calculation that a non-zero contribution  arises
only at order $1/q$.  The variables ($\rho, G(n), w(n)$) must be expanded to
order $1/q^2$ to consistently extract this contribution, because the
Hamiltonian contains an explicit factor of $q$.   Indeed, as we will see
below, for the $\rho_0=0$ solution, the second order expansion of these
variables does not in fact give a non-zero contribution, however they do 
contribute to
the energy for the $\rho_0=\lambda$ solution.
Then we find the solution to second order:
\begin{eqnarray}
G_{2}(n)&=&\alpha_{2}n^{5}+\beta_{2}n^{4}+\delta_{2}n^{3}+\gamma_{2}n^{2}+\xi_{2}n+\epsilon_{2},
\\
\alpha_{2}&=&\frac{\alpha}{20},\,\,\beta_{2}=\frac{\beta}{12},\,\,\delta_{2}=\frac{-\alpha+2\delta}{12}\nonumber\\
\xi_{2}&=&-\frac{\alpha
N^{4}}{20}+\frac{N^{3}}{12}(3\alpha-\beta)+\frac{N^{2}}{12}(-5\alpha+4\beta-2\delta)\\
&+&\frac{N}{12}(3\alpha-5\beta+6\delta-6\gamma)+\frac{1}{6}(\beta-3\delta+6\gamma)\nonumber\\
\gamma_{2}&=&\frac{-\beta+6\gamma}{12},\,\,\epsilon_{2}=\xi_{2}-\frac{\alpha-5\delta+15\gamma}{30},
\\
\alpha&=&\frac{\lambda^{2}\alpha_{1}}{2},\,\,\beta=\frac{\lambda^{2}}{4}(2\beta_{1}-1),\,\,\delta=\frac{\lambda}{2}(2\rho_{1}+\lambda\delta_{1}-\lambda)\nonumber\\
\gamma&=&\lambda\rho_{1}-\rho_{1}^{2}+\lambda\rho_{2}-\frac{\lambda^{2}}{4}-2\rho_{2}\rho_{0}+\frac{\lambda^{2}\gamma_{1}}{2}\\
\rho_{2}&=&\frac{-(N^{2}-1)(\lambda^{3}+2(2\rho_{0}-\lambda))}{96}.
\end{eqnarray}
We have implicitly assumed that $q\gg \lambda$, $q\gg
\rho_{i}$ and $q\gg G_{i}(n)$ for every $i$ (or more precisely that these
variables are of order of $q^{0}$). However, we find that the $G_{i}(n)$ are
polynomial in $n$ where $n=0,\cdots ,N$. Thus if $N$ becomes large,
$G_{i}(n)$ would also become large.  Therefore the condition $q\gg 1$ is
not adequate for the perturbative expansion to converge. We see that
$G_{1}(n)$  and $G_{2}(n)$ are respectively third and fifth degree
polynomials. Thus our perturbation series would not be valid if $N^{3}\gg q$
or $N^{5}\gg q^{2}$.   If we replace
\begin{equation}
G(n)=\sum_{i=0}^{\infty} \frac{G_{i}(n)}{q^{i}}.
\end{equation}
in equation (\ref{numeric}) and expand to  order $1/q^{i}$, then we
obtain a recurrence equation relating the discrete
Laplacian of $G_{i}(n)$ to $G_{j}(n)$ with $j<i$. Assuming that the solution
for the $G_{j}(n)$ with $j<i$ is a polynomial in $n$, the solution for
$G_{i}(n)$ is a polynomial for degree two higher.   Then knowing that
$G_1(n)$ is a polynomial of order 3, by induction
we see that $G_{i}(n)$ is a polynomial of the degree $2i+1$. Thus we see
that the true perturbative parameter is actually
is $n^2/q$. Since $n=0,\cdots ,N$, we take the strongest condition,
$N^2/q\ll 1$.  This condition might actually not be necessary, since we in
fact impose the boundary condition that $G(N-1)=0$, and indeed our numerical
analysis agrees with the perturbative analysis even for $N^2/q\sim N$.

The Hamiltonian is made up of the kinetic and potential
energy. The potential energy actually contributes at order $1/q^2$,
which we will neglect. The kinetic energy contains terms of order
$q$. These are in principle the dominant terms for large $q$. They
have a constant energy density, that is the energy associated to
each state$\mid n\rangle\langle n\mid$. In total the trace gives
\begin{equation}
H_{kin.}(q)=2\Xi N\rho_{0}^{2}q.
\end{equation}
At order zero, the kinetic energy density is
linear (in $n$), but the total after the trace is zero:
\begin{equation}
H_{kin.}(q^{0})=2\Xi Tr\left(-\lambda\rho_{0}
\sum_{n=0}^{N-1}\left(n-\frac{2\rho_{1}}{\lambda}\right)\mid
n\rangle\langle n\mid\right)=0.\label{lin}
\end{equation}
For the order $1/q$ contribution we need $w(n)^{2}$ to
order $1/q^2$.
\begin{equation}
(w(n))^{2}\simeq
\rho_{0}^{2}-\frac{\lambda\rho_{0}}{q}\left(n+1-\frac{2\rho_{1}}{\lambda}\right)+\frac{1}{q^2}\left(\frac{\lambda^{2}}{4}\left(n+1-\frac{2\rho_{1}}{\lambda}\right)^{2}+2\rho_{0}\left(\rho_{2}-\frac{\lambda
G_{1}(n)}{2}\right)\right).
\end{equation}
The kinetic energy then is
\begin{equation}
H_{kin.}(1/q)=\frac{2\Xi }{q}Tr\left(
\sum_{n=0}^{N-1}\left(\frac{\lambda^{2}}{4}(n-\frac{2\rho_{1}}{\lambda})^{2}+2\rho_{0}\rho_{2}\right)\mid
n\rangle\langle n\mid\right)\label{quad}
\end{equation}
which gives
\begin{equation}
H_{kin.}(1/q)=\frac{2\Xi
}{q}\left(\frac{\lambda^{2}N}{48}(N^{2}-1)+2\rho_{0}\rho_{2}N\right).
\end{equation}
Hence the energy
up to the $1/q$  is
\begin{equation}
H\simeq 2\Xi N\rho_{0}^{2}q+\frac{2\Xi
}{q}\left(\frac{\lambda^{2}N}{48}(N^{2}-1)+2\rho_{0}\rho_{2}N\right).
\end{equation}
The two solutions (for $\rho_0$) behave quite differently for
$\rho_{0}=\lambda$ the energy diverges as $q$ becomes large while for
$\rho_{0}=0$ it vanishes.
\begin{equation}
H_{\rho_{0}=\lambda}\simeq 2\Xi N\lambda^{2}q+\frac{2\Xi
}{q}\left(\frac{\lambda^{2}N}{48}(N^{2}-1)+2\lambda\rho_{2}N\right)
\end{equation}
\begin{equation}
H_{\rho_{0}=0}\simeq \frac{\Xi\lambda^{2}N(N^{2}-1) }{24q}
\end{equation}
These solutions correspond to an annulus of large radius and small
(relatively) thickness that oscillate in time.  We can see for the operator
of the radius squared (which is diagonal), from equation
(\ref{R2}), the dominant part of the coefficient (hence eigenvalue of $R^2$)
is proportional to
$2q$ for every state, in the  large $q$ limit. The next
dominant term is proportional to the square root of $q$ and is an
oscillatory term (the frequency  for $\rho_{0}=\lambda$ is of order $q^0$ 
and it is weakly $n$ dependent,
and for $\rho_{0}=0$ it is of order $q^{-1}$ ).
The next important
term is proportional to $2n+1$.  As $q\gg \sqrt{q}\gg 1$ the radius is very
large
and oscillates with an amplitude that is relatively much smaller than the
radius. Thus from afar the droplet looks like an thin annulus that undergoes
a nontrivial oscillation.

\subsubsection{Small quasi-hole solution}

The small quasi-hole solution, for ($q\ll 1$), can be obtained by a
perturbation on the solution found in \cite{Isabeau}.  Take $D$ at first
order perturbation to have the form
\begin{equation}
D=D'+D_{p},\,\,D_{0}=D_{0}'+D_{0p},\,\,\Psi=\Psi'+\Psi_{p},
\end{equation}
where the primed variables are solutions found in \cite{Isabeau} and the
variables subscripted $p$ are the perturbations.  To first order the Gauss
law gives
\begin{eqnarray}
&0&=\lefteqn{[D_{p},[D_{0}',D'^{\dagger}]]+[D',[D_{0p},D'^{\dagger}]]+[D',[D_{0}',D_{p}^{\dagger}]]+[D_{p}^{\dagger},[D_{0}',D']]}\nonumber\\
&+&[D'^{\dagger},[D_{0p},D']]+[D'^{\dagger},[D_{0}',D_{p}]]+
\lambda([D_{p},D'^{\dagger}]+[D',D_{p}^{\dagger}])+\Psi_{p}\Psi'^{\dagger}+\Psi'\Psi_{p}^{\dagger}
\end{eqnarray}
while the Amp\`{e}re law gives
\begin{eqnarray}
\lefteqn{[D_{0p},[D_{0}',D']]+[D_{0}',[D_{0p},D']]+[D_{0}',[D_{0}',D_{p}]]+[D_{p},[D',D'^{\dagger}]]}\nonumber\\
&+&[D',[D_{p},D'^{\dagger}]]+[D',[D',D_{p}^{\dagger}]]=\lambda[D_{0p},D']+\lambda[D_{0}',D_{p}]
\end{eqnarray}
and the constraint on $\Psi$ gives
\begin{equation}
D_{0p}\Psi'+D_{0}'\Psi_{p}=0\label{equation de psi perturbatif}.
\end{equation}
We assume the perturbation takes the form
\begin{equation}
D_{p}=\sqrt{q}e^{\frac{i\rho\sqrt{\kappa} t}{\sqrt{\theta}}}\mid
N-1\rangle\langle 0\mid,\,\,D_{0p}=0,\,\,\Psi_{p}=0\label{P} .
\end{equation}
The ansatz in \cite{Isabeau} has the form (gauge equivalent to)
\begin{equation}
D'=\sum_{n=0}^{N-2}\sqrt{G(n)}e^{iw(n)t}\mid n\rangle\langle n+1\mid,\,\,
A_{0}=0
\end{equation}
where $G(n)$ and $w(n)$ are explicitly calculated (numerically) in
\cite{Isabeau}.
Then the Gauss law is automatically satisfied using
\begin{equation}
\left[\mid N-1\rangle\langle 0\mid,\sum_{n=0}^{N-2} f(n)\mid
n+1\rangle\langle n\mid\right]=\left[\mid 0\rangle\langle
N-1\mid,\sum_{n=0}^{N-2} f(n)\mid n\rangle\langle n+1\mid\right]=0.
\end{equation}
The Amp\`{e}re law becomes
\begin{equation}
[D_{0}',[D_{0}',D_{p}]]+[D_{p},[D',D'^{\dagger}]]=\lambda[D_{0}',D_{p}]
\end{equation}
with $[D',D'^{\dagger}]$ given by \ref{DDdag}.  This  implies the
same condition as before (removing an overall factor $D_{p}$)
(\ref{numeric}) however only for $n=N-1$
\begin{equation}
\rho^{2}D_{p}+(G(0)+G(N-2)) D_{p}=\lambda \rho D_{p}.
\end{equation}
For other values of $n$ the condition is automatically satisfied.
Then $\rho$ is given by
\begin{equation}
\rho=\frac{\lambda\pm\sqrt{\lambda^{2}-4(G(0)+G(N-2))}}{2}.
\end{equation}
As this expression involves a square root, we see that there is no solution 
for small $\lambda$, as we will confirm below, in the numerical analysis.  
This is a gap in the spectrum of allowed $q$ for small $\lambda$.
The change in energy is given by
\begin{equation}
\Delta H =-2\Xi
Tr([D_{0}',D_{p}^{\dagger}][D_{0}',D']+[D_{0}',D'^{\dagger}][D_{0}',D_{p}]+[D_{0}',D_{p}^{\dagger}][D_{0}',D_{p}])
\end{equation}
which gives
\begin{equation}
\Delta H =-2\Xi Tr(-\rho
D_{p}^{\dagger}[D_{0}',D']+\rho[D_{0}',D'^{\dagger}]D_{p}-\rho^{2}D_{p}^{\dagger}D_{p})\label{C}
\end{equation}
and using that the two first terms vanish, yields
\begin{equation}
\Delta H =2\Xi \rho^{2}q\label{CT}.
\end{equation}

\subsection{Numerical analysis of the quasi-hole solution}

In this section, we will find numerical solutions of
equation (\ref{numeric}).  As our solutions are built upon the solutions
found in \cite{Isabeau}, we first give the numerical analysis of the
equations considered there.  We take this opportunity to correct certain
errors that have appeared in \cite{Isabeau}.  Equation (5.6) in
\cite{Isabeau} is incorrect, the kinetic energy is not symmetric about
$M=N/2$.  The correct equation is:
\begin{equation}
T =  2 \Xi \Bigg(  \Big\{ \sum_{n=0}^{N-2}  2g^2 u_n - u_n
\nabla^2u_n \Big\} + g^2(N^2-N-2NM) \Bigg) \label{num}
\end{equation}
Correspondingly, figure (5) in \cite{Isabeau} is also not correct.  In
addition, there seems to be an inconsistency between  figure (5) and figure
(7) in \cite{Isabeau}, the kinetic energy appears to be greater than the
total energy.  We give corrected figures, figure  \ref{energie cin fig
Isabeau} and figure \ref{energie
pot fig Isabeau} here (not for the same values of the parameters).
\DOUBLEFIGURE{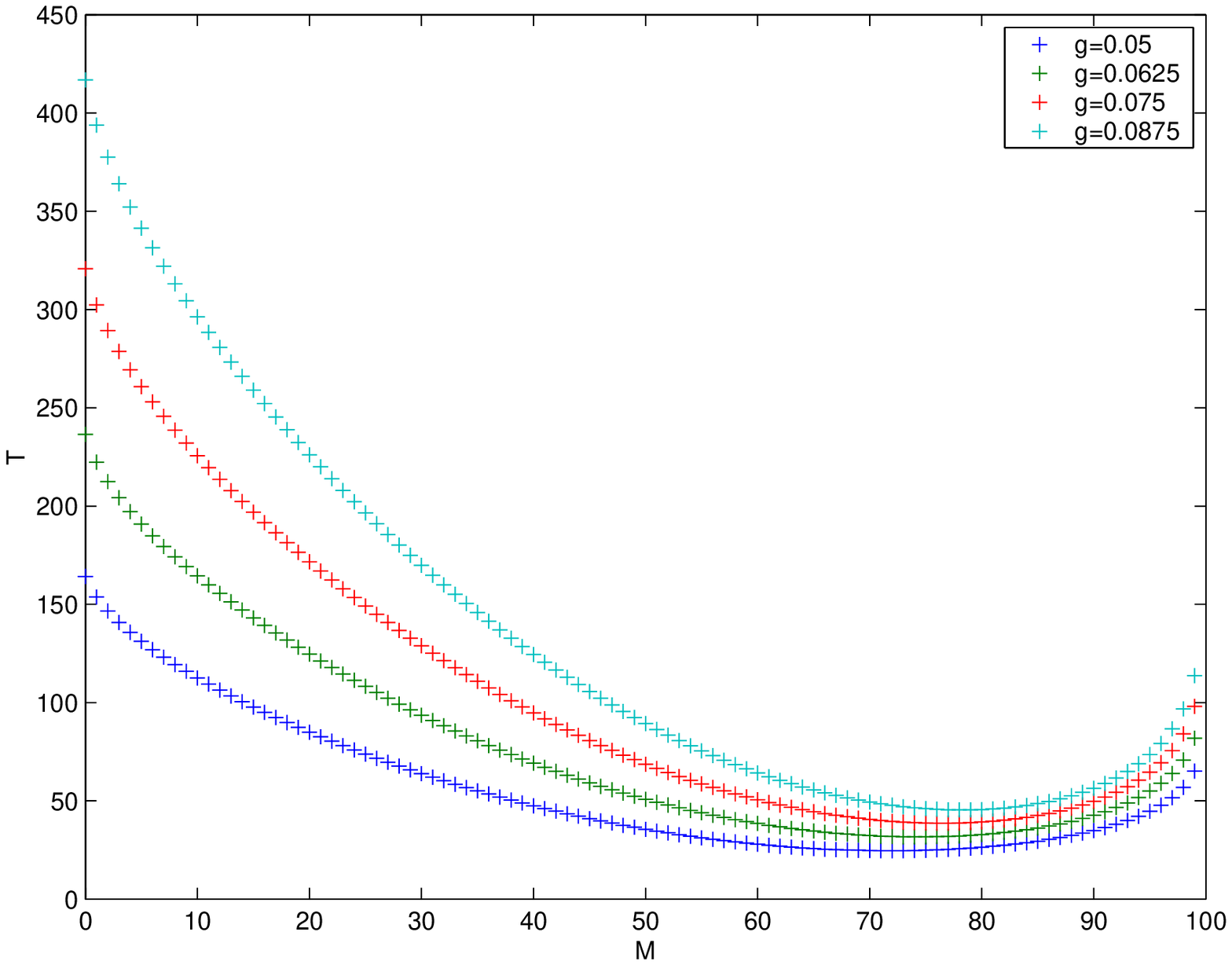,scale=0.4,
angle=0}{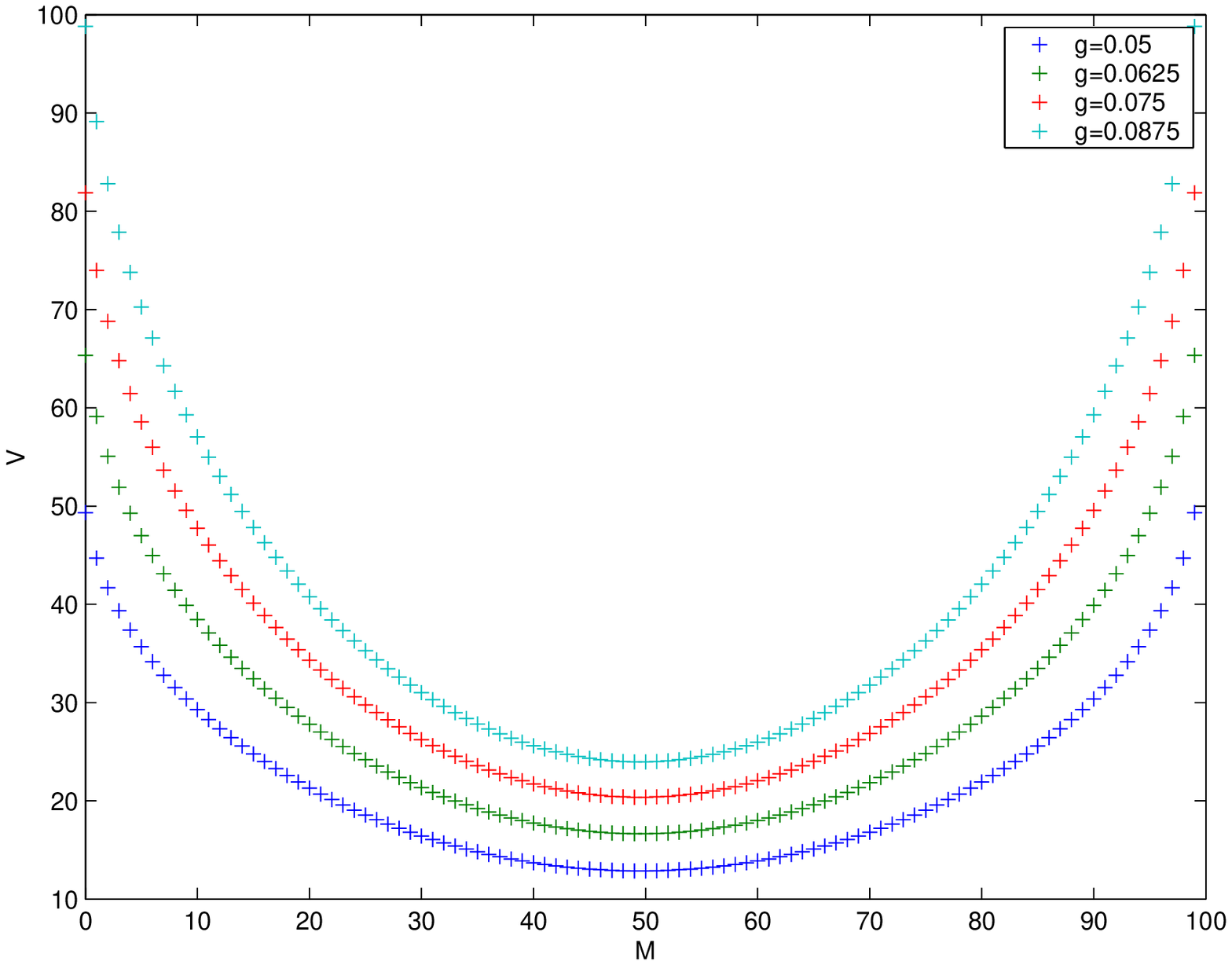,scale=0.4, angle=0}{Kinetic energy as a
function of M for N=100 for (corrected) solution in
\cite{Isabeau}\label{energie cin fig Isabeau}}{Potential energy as a
function of M for N=100 for solution in \cite{Isabeau}\label{energie
pot fig Isabeau}} These are representative solutions for the system of
equations studied in \cite{Isabeau}.  Our further numerical analysis
concerns solutions built upon these.
We ultimately use  the Newton method \cite{nm} to solve the difference equations.
Finite difference equations on a finite set of variables/parameters and
boundary conditions, after iterating recursively and removing dependent
variables,  simply become a system of extremely complicated algebraic
equations in a much reduced set of variables.    In our case, we reduce the
system to two equations in three variables, which we take to be $q$, $\rho$
and $G(N-2)$.  We solve this resultant system by the Newton method.  We can
constrain our search a little by noting that $q+G(n)\geq 0$ since it appears
under a square root in all the expressions for $D$.  For the case $n=N-1$,
if we add $2q$ to the equation (\ref{numeric})  then we get the constraint
\begin{equation}
\rho\in\left[\frac{\lambda- \sqrt{\lambda^{2}+8q}}{2},\frac{\lambda+
\sqrt{\lambda^{2}+8q}}{2}\right]
\end{equation}

Numerically, we easily find the two solutions that we have determined
analytically, for $q\gg 1$.
\DOUBLEFIGURE{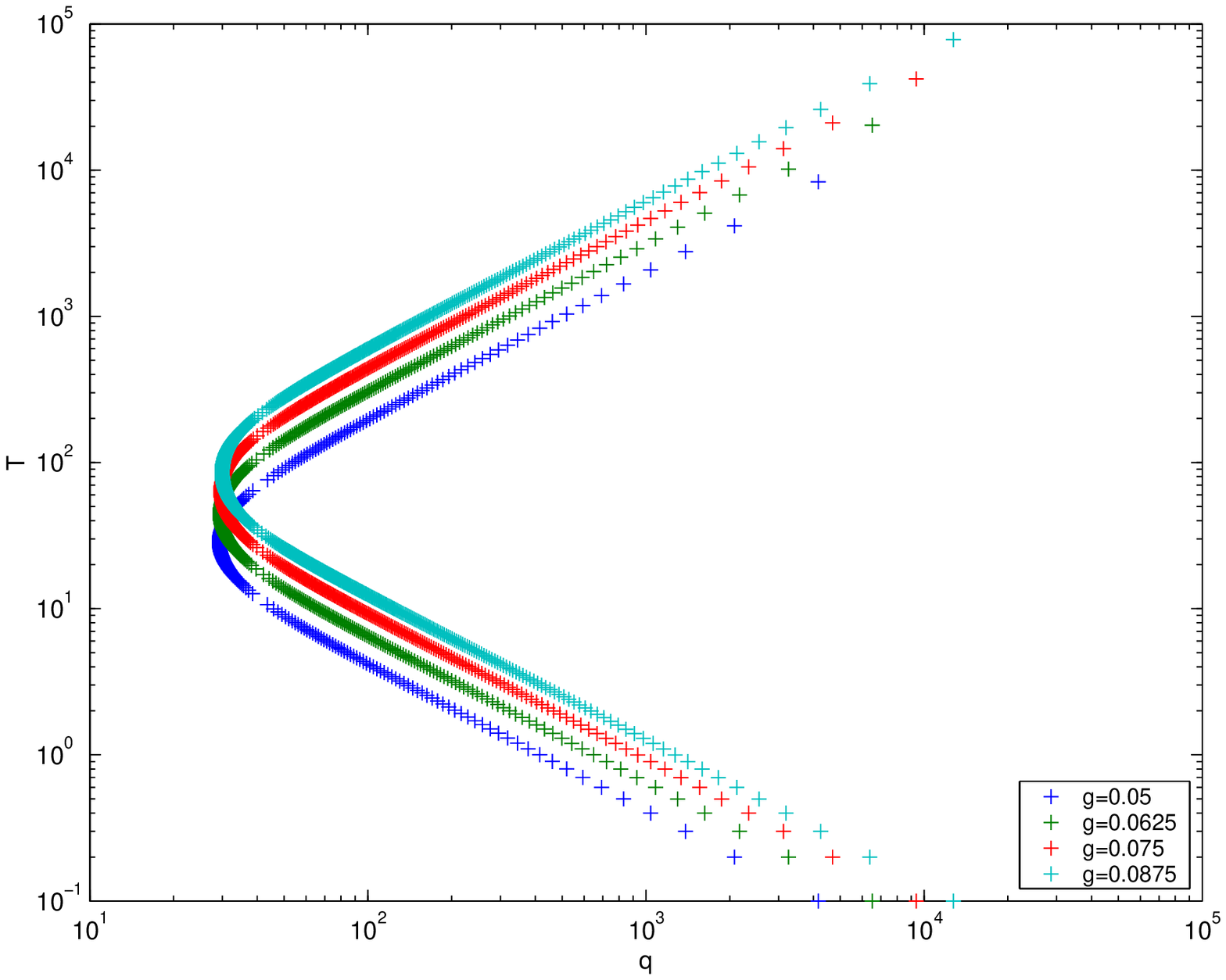,scale=0.4,
angle=0}{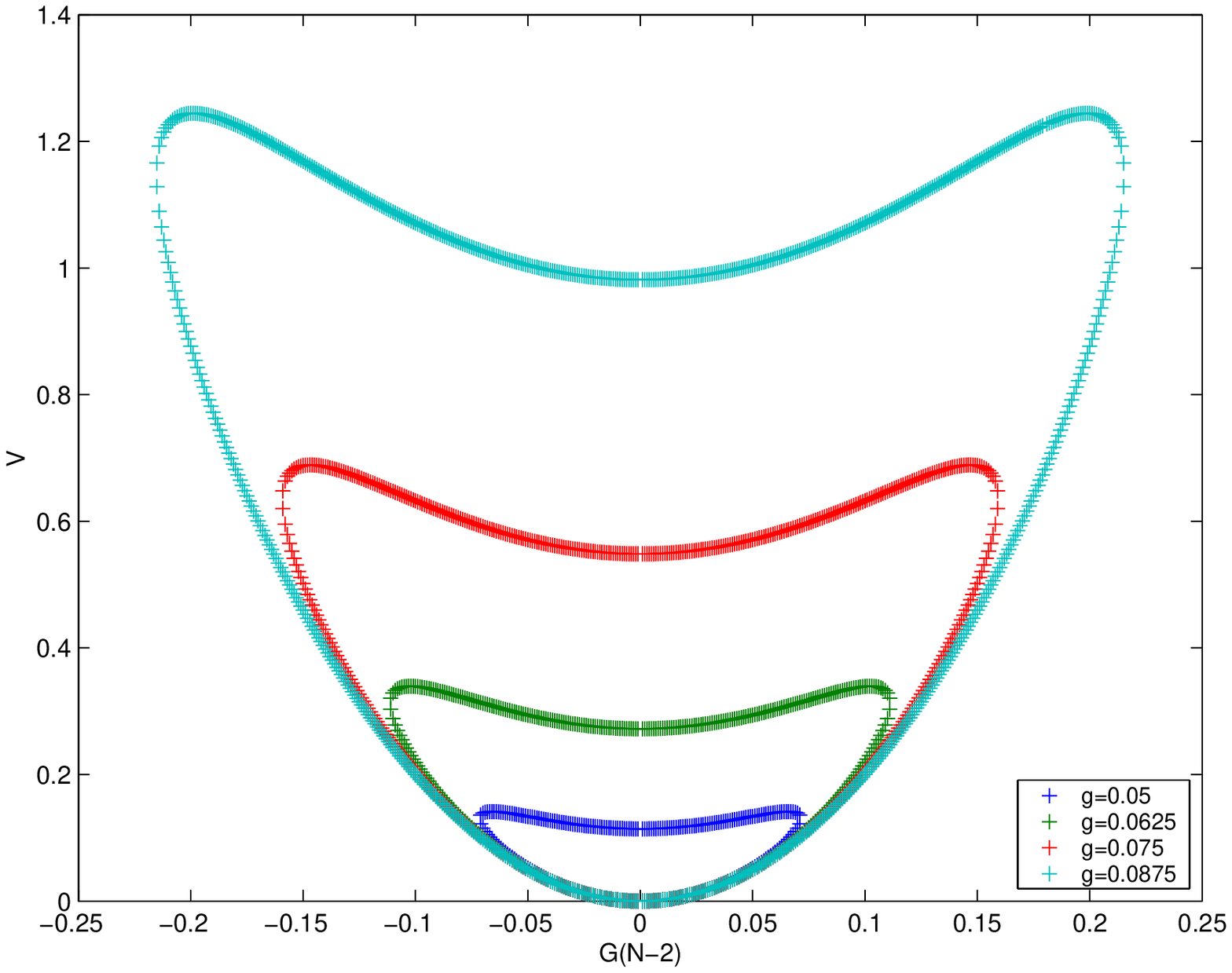,scale=0.4, angle=0}{Kinetic energy as a
function of $q$ for $N=100$ and for 4 values of $g=\lambda /2$ in units of
$\Xi$.\label{energie cin fig}}{Potential energy as a function of $G(N-2)$
for $N=100$ and $g=\lambda /2$ in unit of $\Xi$.\label{energie pot
fig}}
We see in figure (\ref{energie cin fig}) that the kinetic energy either
diverges as $q$ or vanishes as $1/q$   as $q\rightarrow\infty$, depending on
which branch we consider, exactly as we have seen in the section
(\ref{section large q}). For the potential energy, we give a graph as a
function of the value of $G(N-2)$.  The preceding two branches of solutions,
are found in figure (\ref{energie pot fig}), on the bottom part of the
curve, symmetrically on either side of the point $G(N-2)=0$.  This
dependence on $G(N-2)$ can be inferred from the analytic solution in section
(\ref{section large q}).   We can see in figure (\ref{energie
cin fig}) that there exists a region of transition between the two branches
of solutions.  This transition region seems to imply a lower limit on the
permitted value of $q$ for given $\lambda$. The
corresponding transition region in figure (\ref{energie pot
fig}) is the upper part of the solution between the two peaks.  Further
numerical analysis varying $\lambda$ seems to indicate that the transition
region actually extends all the way to $q=0$ for sufficiently large
$\lambda$.  The $q=0$ limit corresponds to the case studied in
\cite{Isabeau}, where it was found that there are $N$ solutions for the
noncommutative droplet.  We find that we recover this multiplicity in the
numerical analysis.  Specifically the curve of the potential energy, figure
(\ref{pot4}) and figure  (\ref{pot3}) obtains $N$ peaks in the transition
region, each of which extends down to the solution found in \cite{Isabeau}
for different choices of $\Psi$, as seen in figures (\ref{kin3}) and
(\ref{kin4}).  (One has to be careful in this limit for $\Psi\nsim
|N-1\rangle$.   One recovers the solution of \cite{Isabeau} for
$(G(N-2-M)+q)\rightarrow 0$ and not simply $q\rightarrow 0$.)  The total
energy is always dominated by the kinetic term and the range of the total
energy is from 0 to $\infty$, however the interesting structure in the
potential energy implies that its contribution to the specific heat could be
most important.
\DOUBLEFIGURE{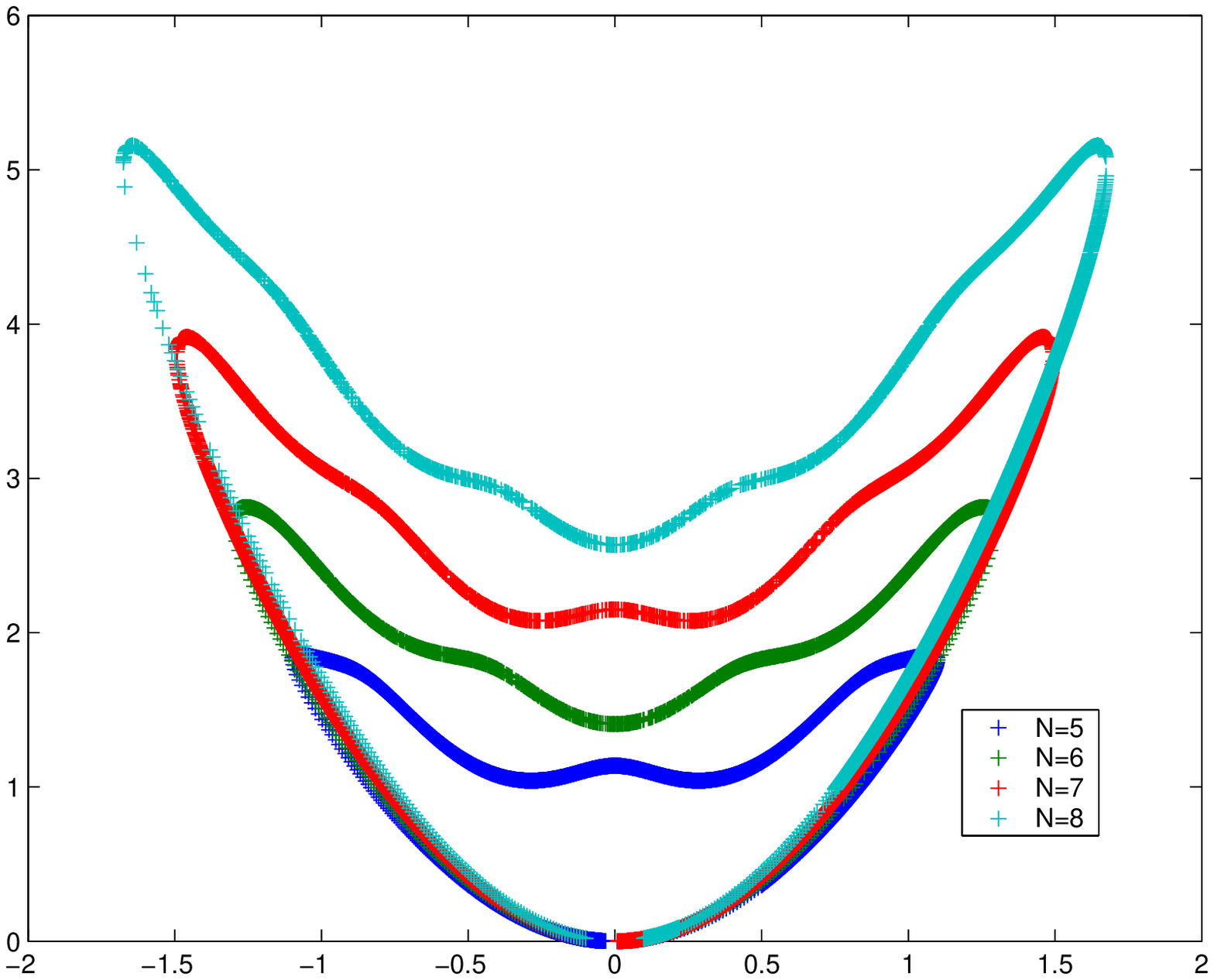,scale=0.4, angle=0}{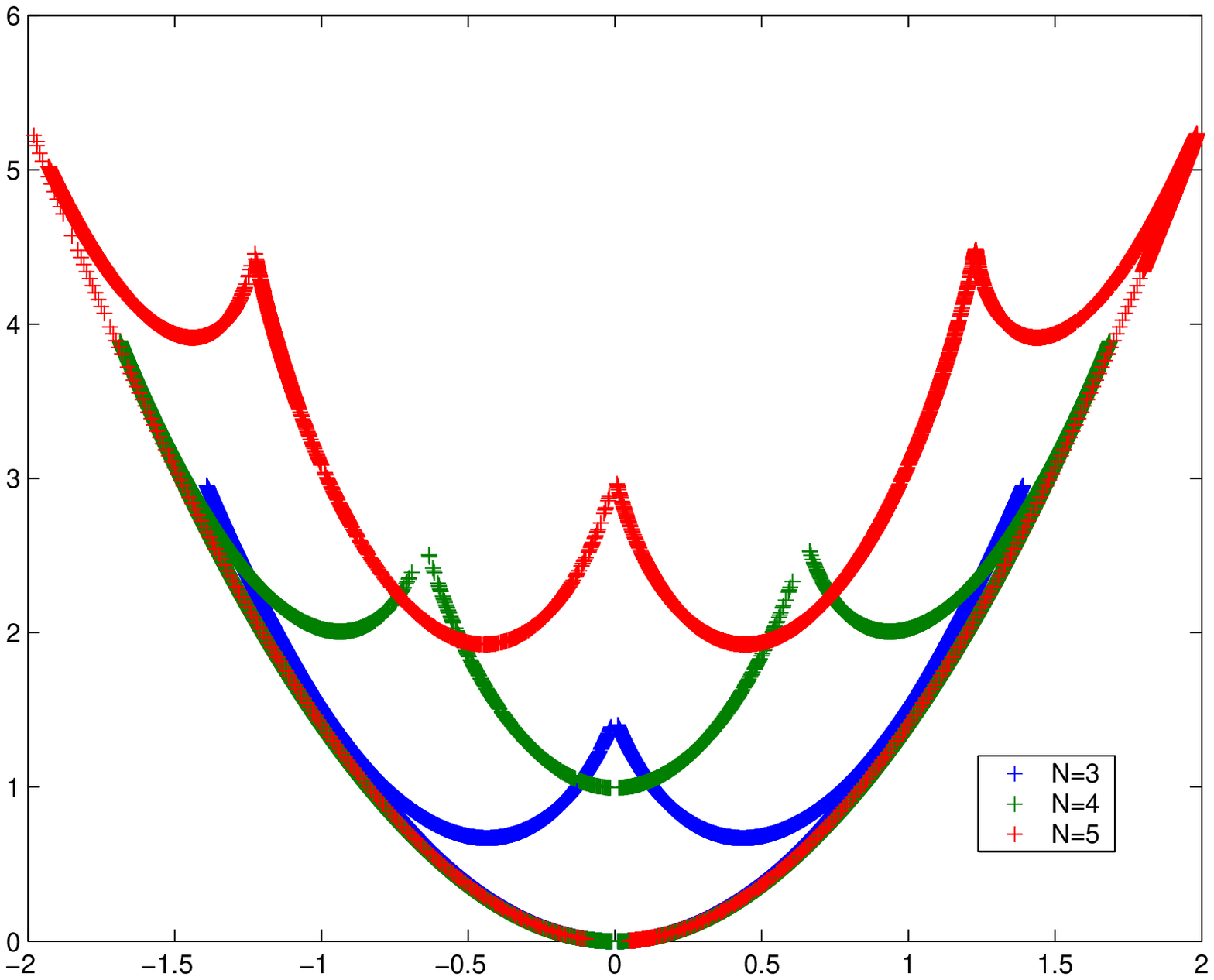,scale=0.4,
angle=0}{Potential energy as a function of $G(N-2)$ for
$g=1$\label{pot4}}{Potential energy as a function of $G(N-2)$ for
$g=1.5$\label{pot3}}
\DOUBLEFIGURE{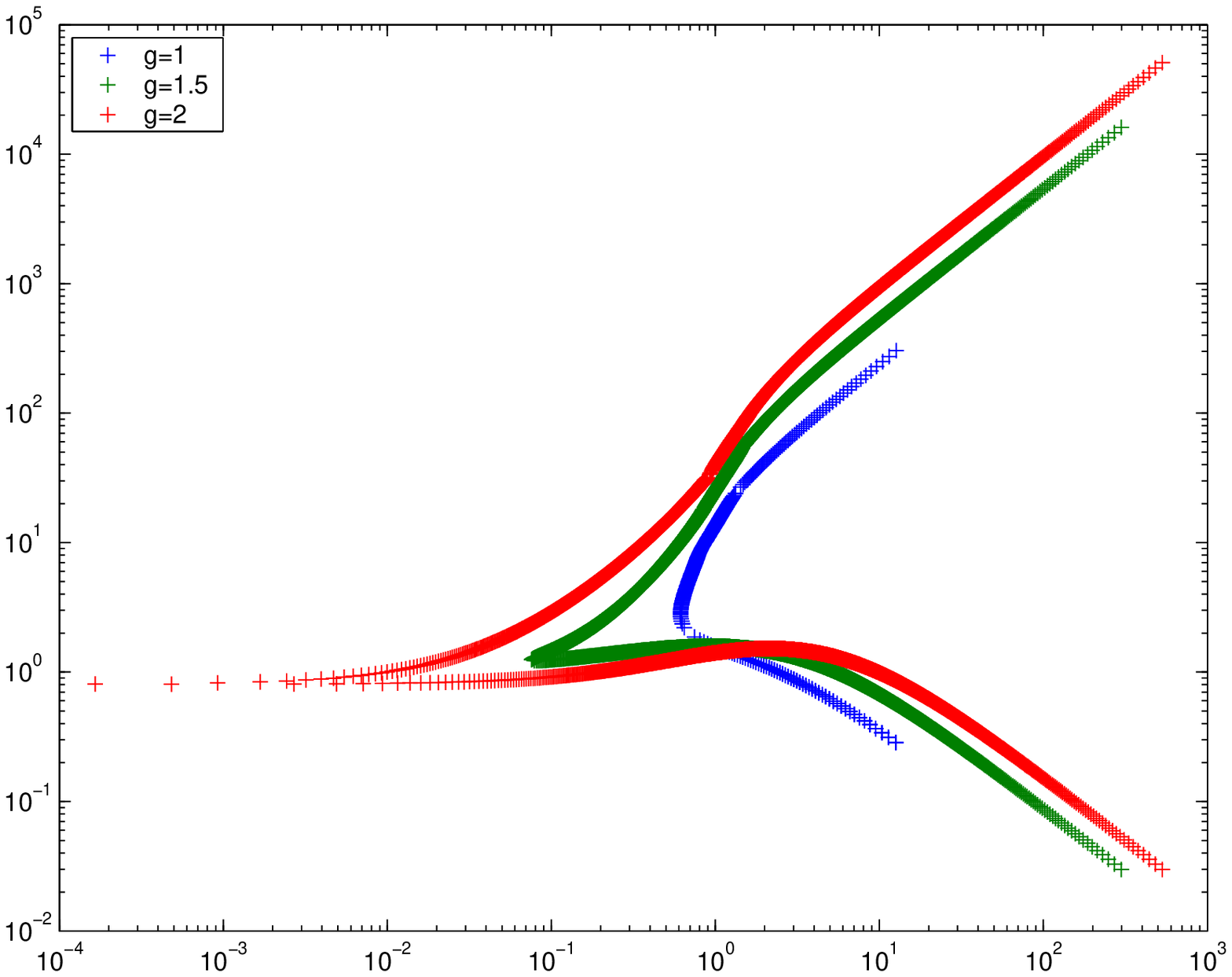,scale=0.4,
angle=0}{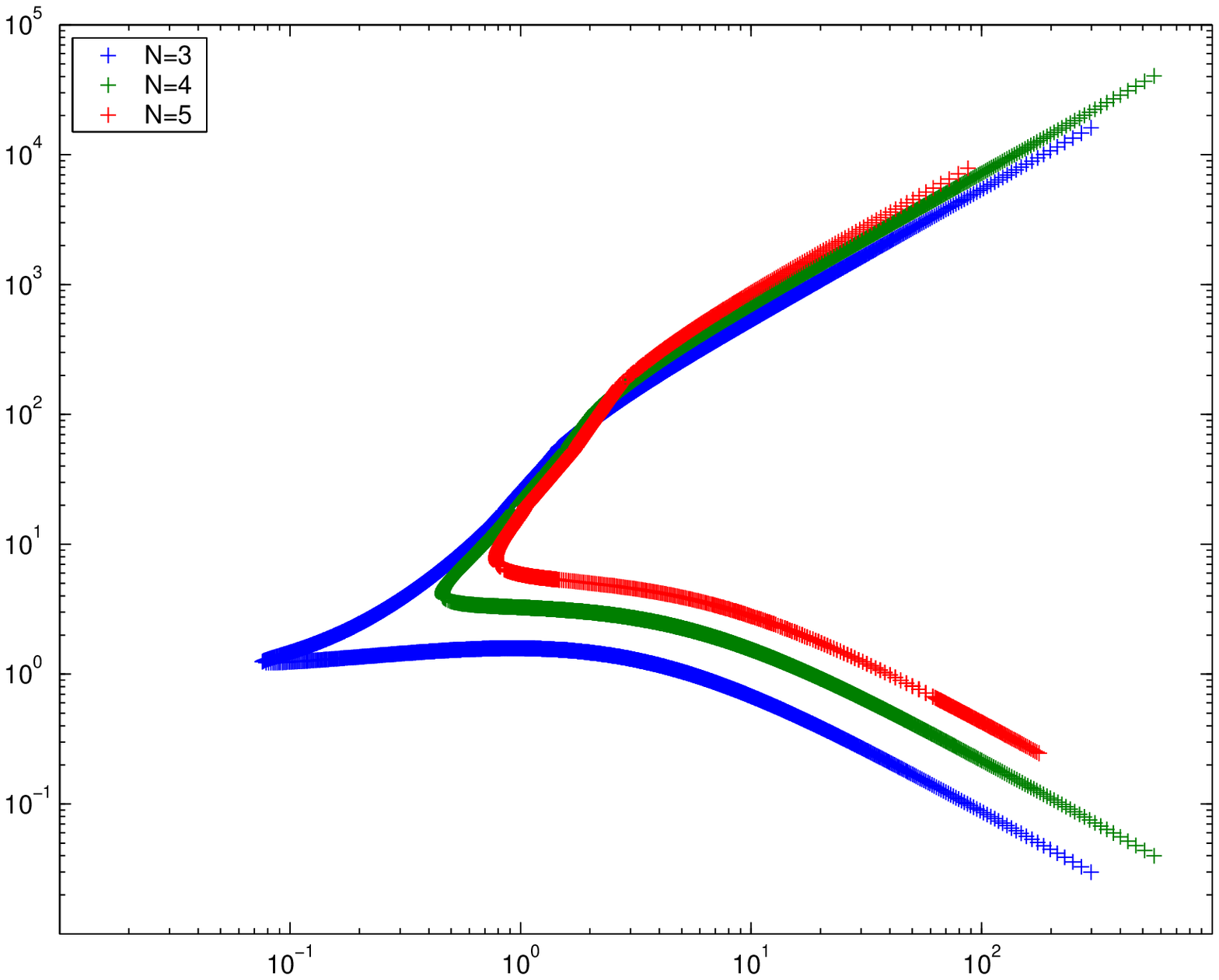,scale=0.4, angle=0}{Kinetic energy as a function of $q$
for $N=3$\label{kin3}}{Kinetic energy as a function of $q$ for
$g=1.5$\label{kin4}}
Comparing our solution to the one found in \cite{Isabeau}, figures
(\ref{energie cin fig Isabeau}) and (\ref{energie pot fig Isabeau}), we find
that the potential energy, figure (\ref{energie pot
fig}), of our
solution is at least one order of magnitude smaller than that found in
\cite{Isabeau} for small $\lambda$ and will approach the potential energy of 
the solutions in \cite{Isabeau} for large $\lambda$ (and small $q$). The 
kinetic energy is, however,  quite different.  In our
solution, the
kinetic energy can take any value since $q$ is a free parameter, while the
solution in
\cite{Isabeau} it is constrained to a discrete set of $N$ values. For
the values of $\lambda$ used in our figures, the kinetic energy of the
solution in \cite{Isabeau} is comparable to the kinetic energy of our
solution in the
transition region.
\DOUBLEFIGURE{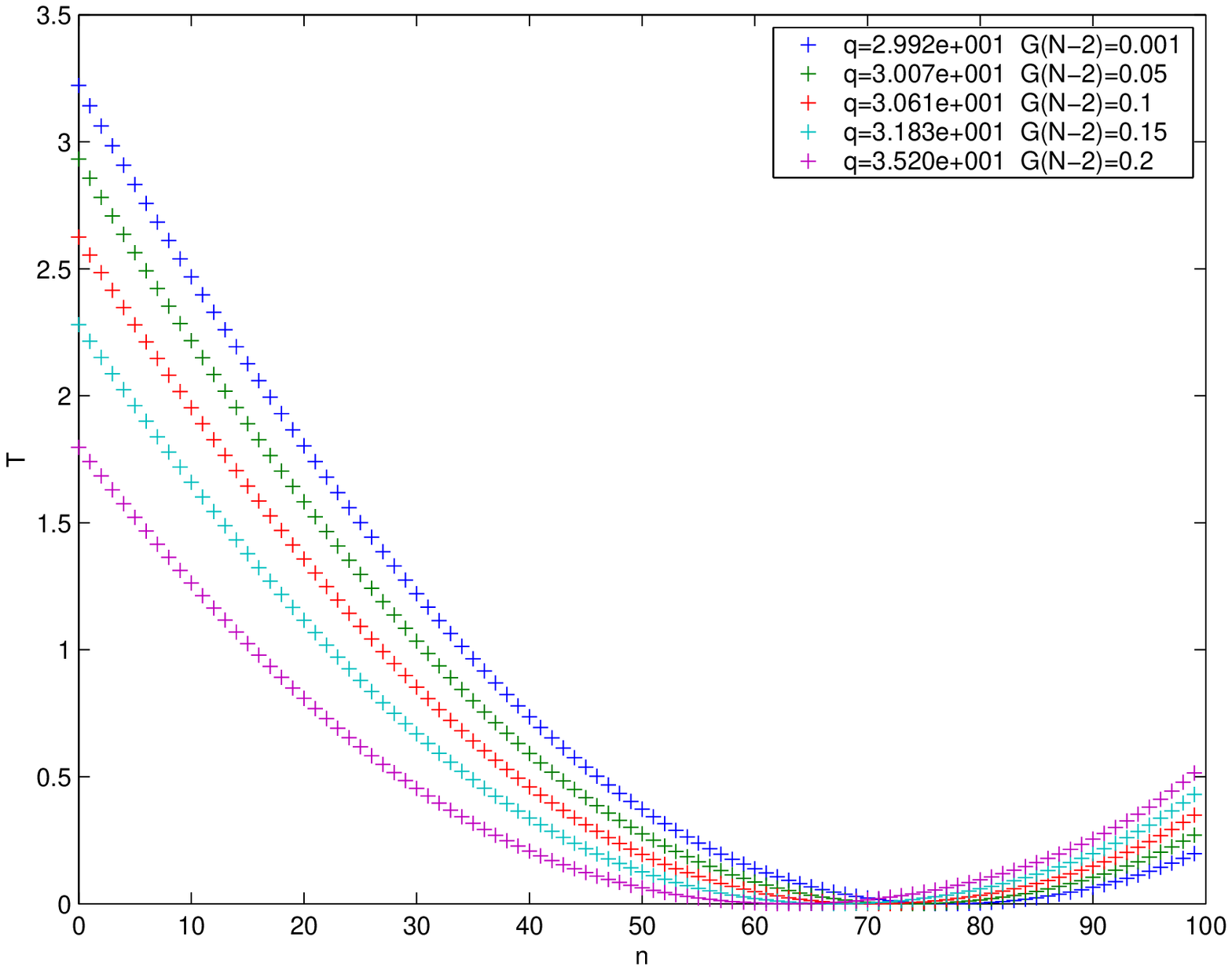,scale=0.4,
angle=0}{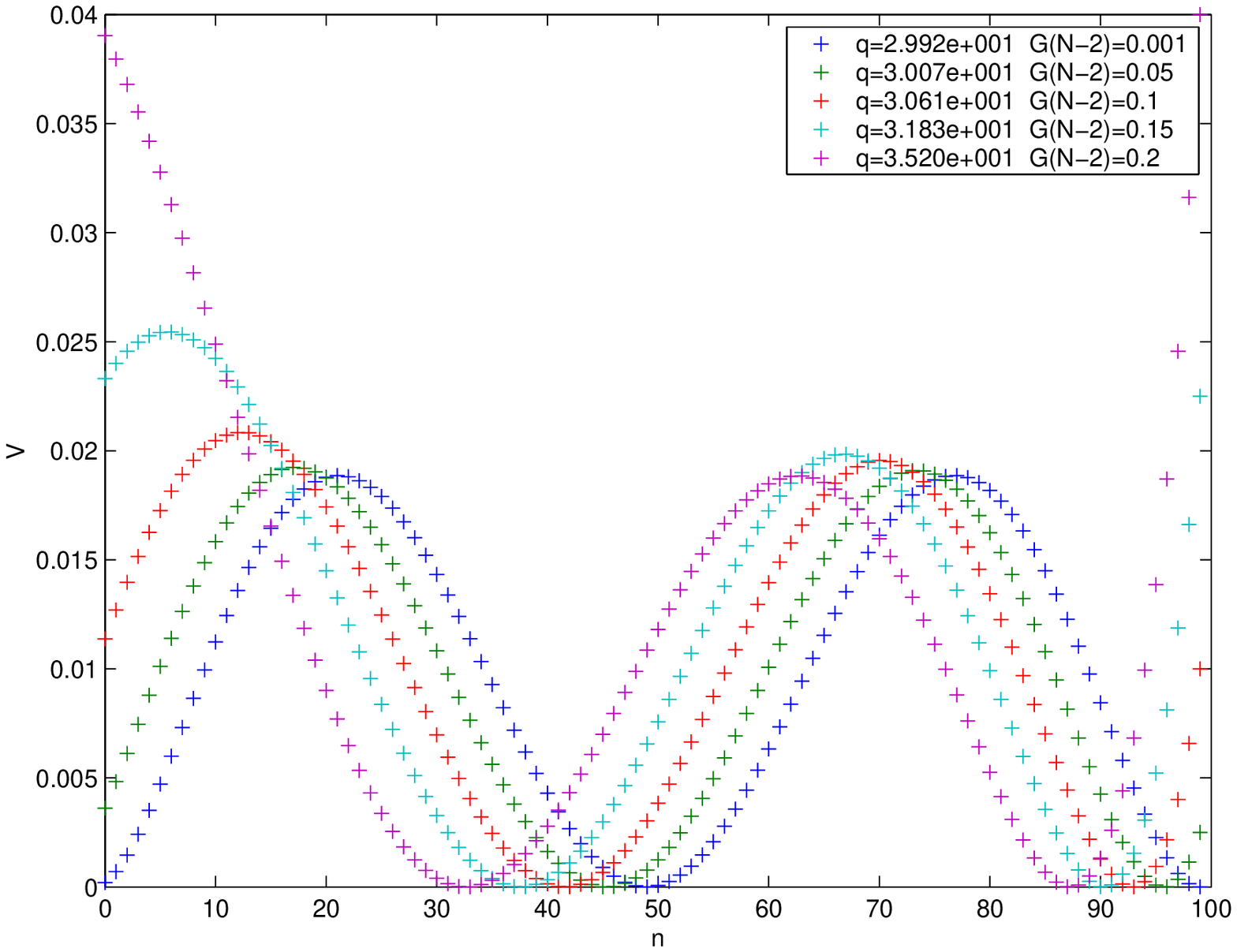,scale=0.4, angle=0}{Kinetic density
energy for $N=100$ and $g=0.0875$\label{energie cin densite
fig}}{Potential density energy for $N=100$ and $g=0.0875$\label{energie
pot densite fig}}

The detail of the transition region is found in the figures (\ref{energie
cin densite fig}), (\ref{energie pot densite fig}) and (\ref{G(n)
fig}) which describes the right half of the transition
region ($G(N-2)>0$) in the figure (\ref{energie pot fig}). The potential
energy density
is seen to translate as we vary $G(N-2)$.   This translation continues for
$G(N-2)<0$
symmetrically with respect to the behavior for $G(N-2)>0$. The potential
energy density is found to concentrate in the bulk of the droplet, away from
the boundaries in the middle of the transition region.  During the transition the kinetic density evolves from a 
linear density
($\rho_{0}=\lambda$) as in equation (\ref{lin}) to a quadratic density
($\rho_{0}=0$) as in equation (\ref{quad}).

Away from the transition  region (i.e. for $q\gg 1$, we do not give a figure
since this region can be computed analytically), the form of the
potential energy density and the $G(n)$'s does not change markedly, apart 
from
their overall size.  The potential energy density is concentrated around the
boundary.

\DOUBLEFIGURE{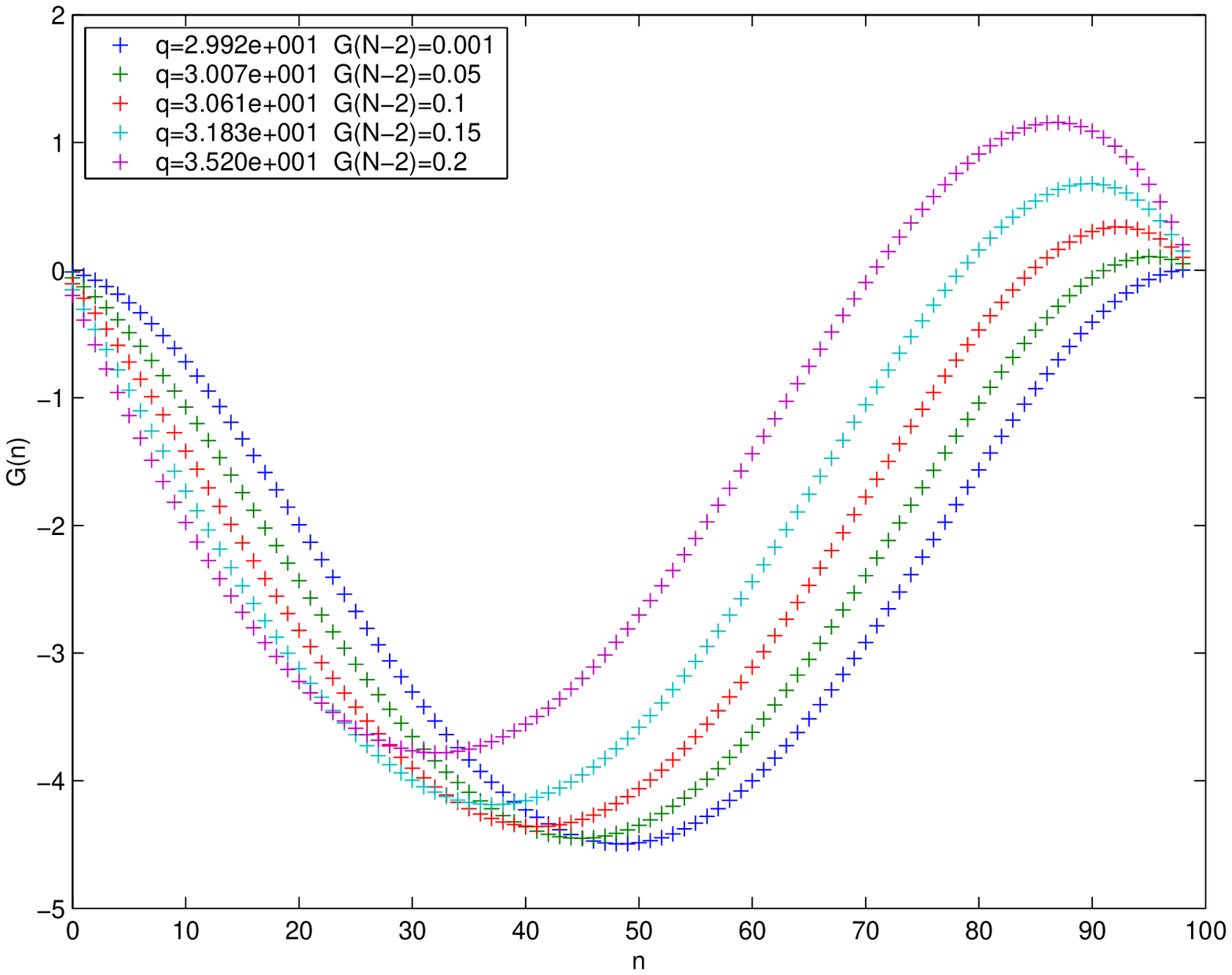,scale=0.4,
angle=0}{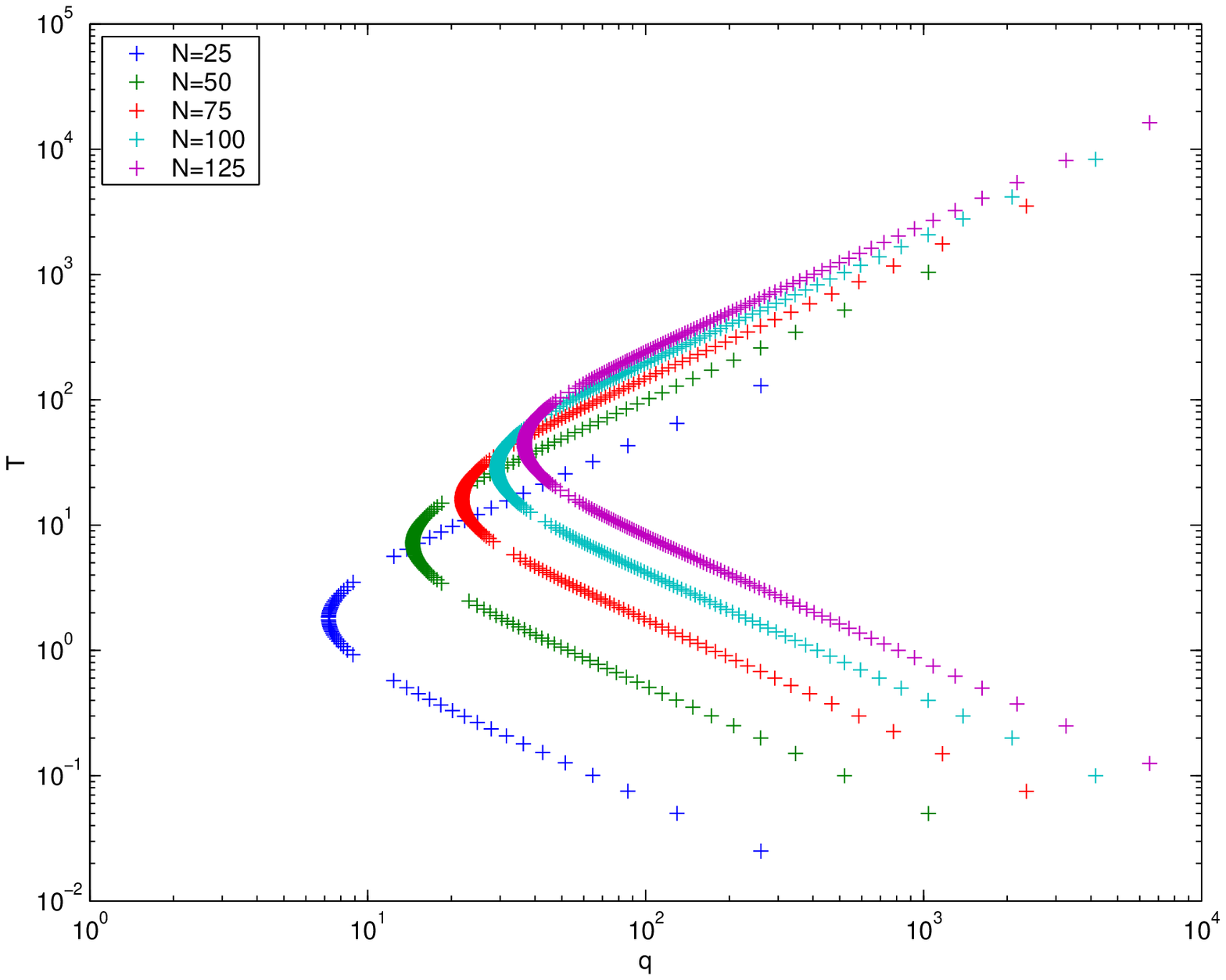,scale=0.4, angle=0}{$G(n)$ for $N=100$ and
$g=0.0875$\label{G(n) fig}}{Kinetic energy as a function of $q$ for
$g=0.05$\label{energie cin en fct N fig}}
Finally, in the figure (\ref{energie cin en fct N fig}), we show the
solutions for various values of $N$.  We see that the minimum value of
$q$ seems to be  growing  linearly with $N$. Thus, assuming that the
trend continues,  as $N\rightarrow\infty$ then
$q_{min.}\rightarrow\infty$, thus the large $q$-hole solutions would not
occur in the infinite plane, at least for small $\lambda$.

\section{Ground state}
The solutions that we have found do not easily allow us to identify the ground state.  For the infinite case, the ground state is given by the solution in  terms of simple annihilation and creation operators, \cite{Susskind, Garnik},
\begin{equation}
a,\,\, a^\dagger, \,\,  [a,a^\dagger ]=1.
\end{equation}
This solution corresponds to a static, quiescent fluid.  However the value of $[D,D^\dagger ]$ is a non-zero constant, which should be considered as the zero point of energy.  The solutions in the finite droplet approach this state arbitrarily closely in the limit $N\rightarrow\infty$ and for large $\lambda $.  We show below in figures (\ref{G(n)}) and (\ref{V}), the plot of $G(n)$ for $M=N-1$ and with $\lambda =2$ and the corresponding vorticity of the fluid.   If we  subtract out the constant background due to the noncommutativity it is evident that the fluid is almost everywhere quiescent with only net vorticity imposed near the boundary.  We take these states as the ground states.
\DOUBLEFIGURE{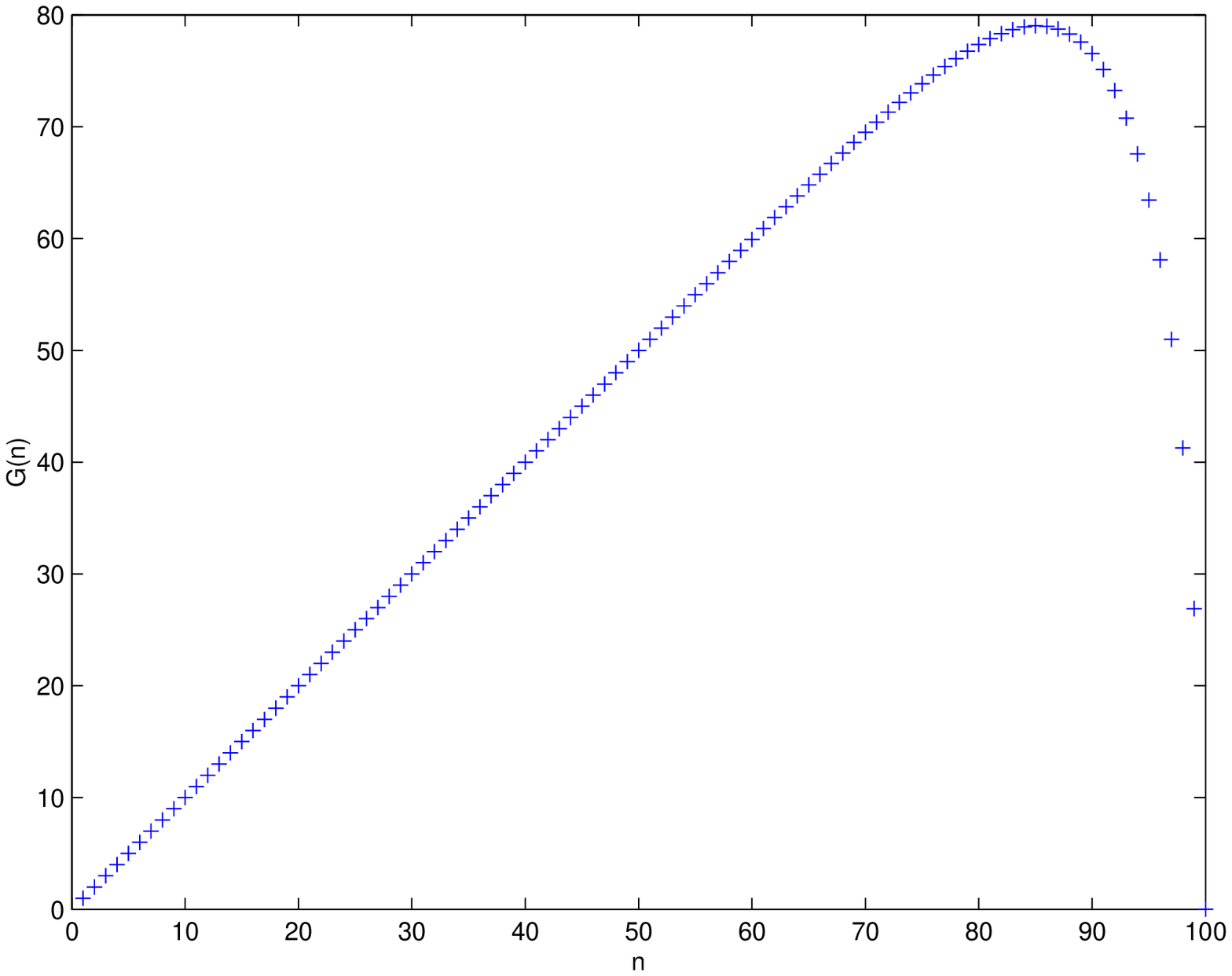, scale=0.4}{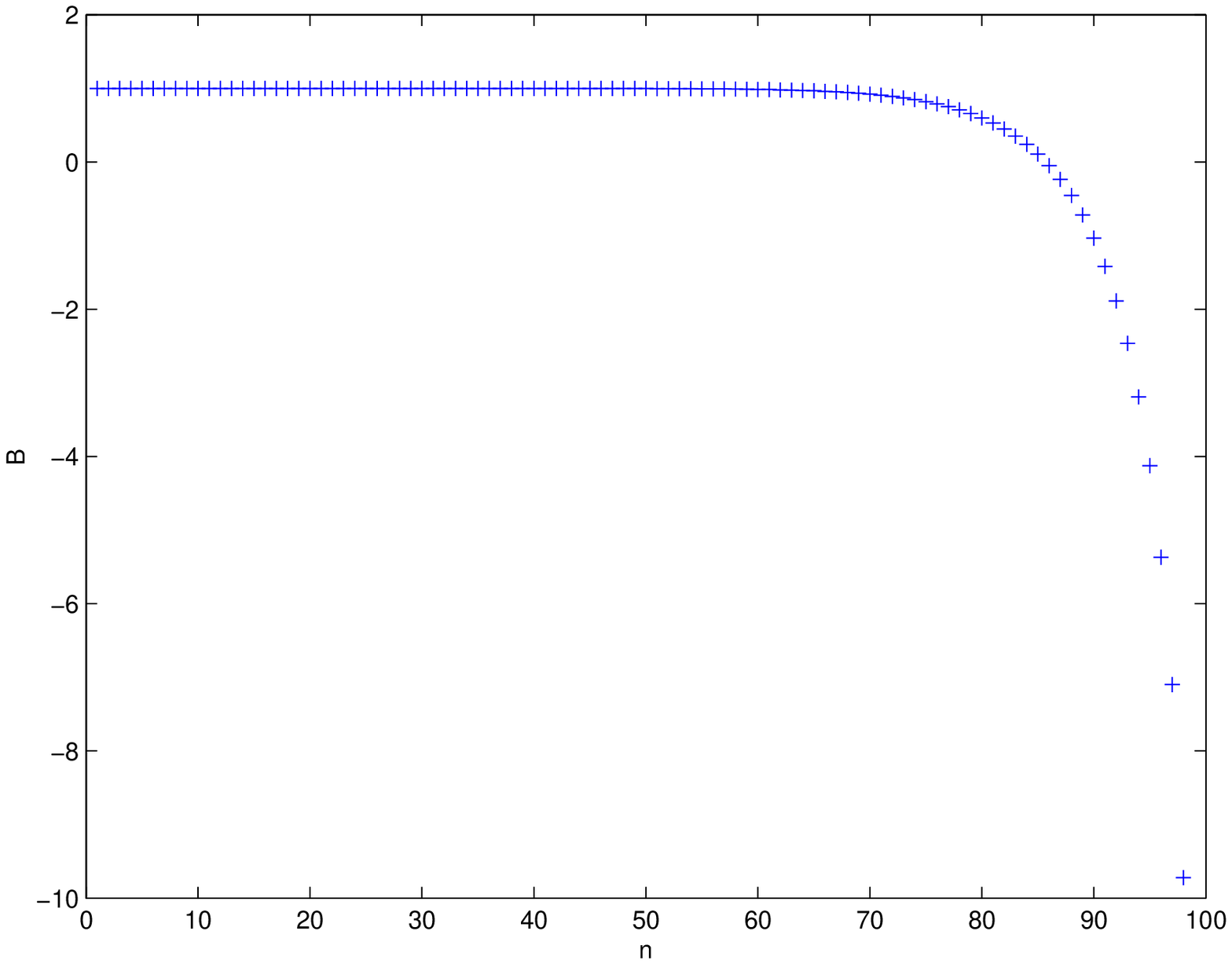, scale=0.4}{\label{G(n)}$G(n)$ for $M=N-1$ with $N=100$}{\label{V}Vorticity  (``magnetic" field $B$ ) for $M=N-1$ with $N=100$ (figure cut-off at $B>-10$)}  
We also show the corresponding kinetic and potential energy in figures (\ref{K}) and (\ref{P}).  The kinetic energy  $T$ is concentrated at the boundary, as is the potential energy $V$ if we subtract off the constant zero point energy.  
\DOUBLEFIGURE{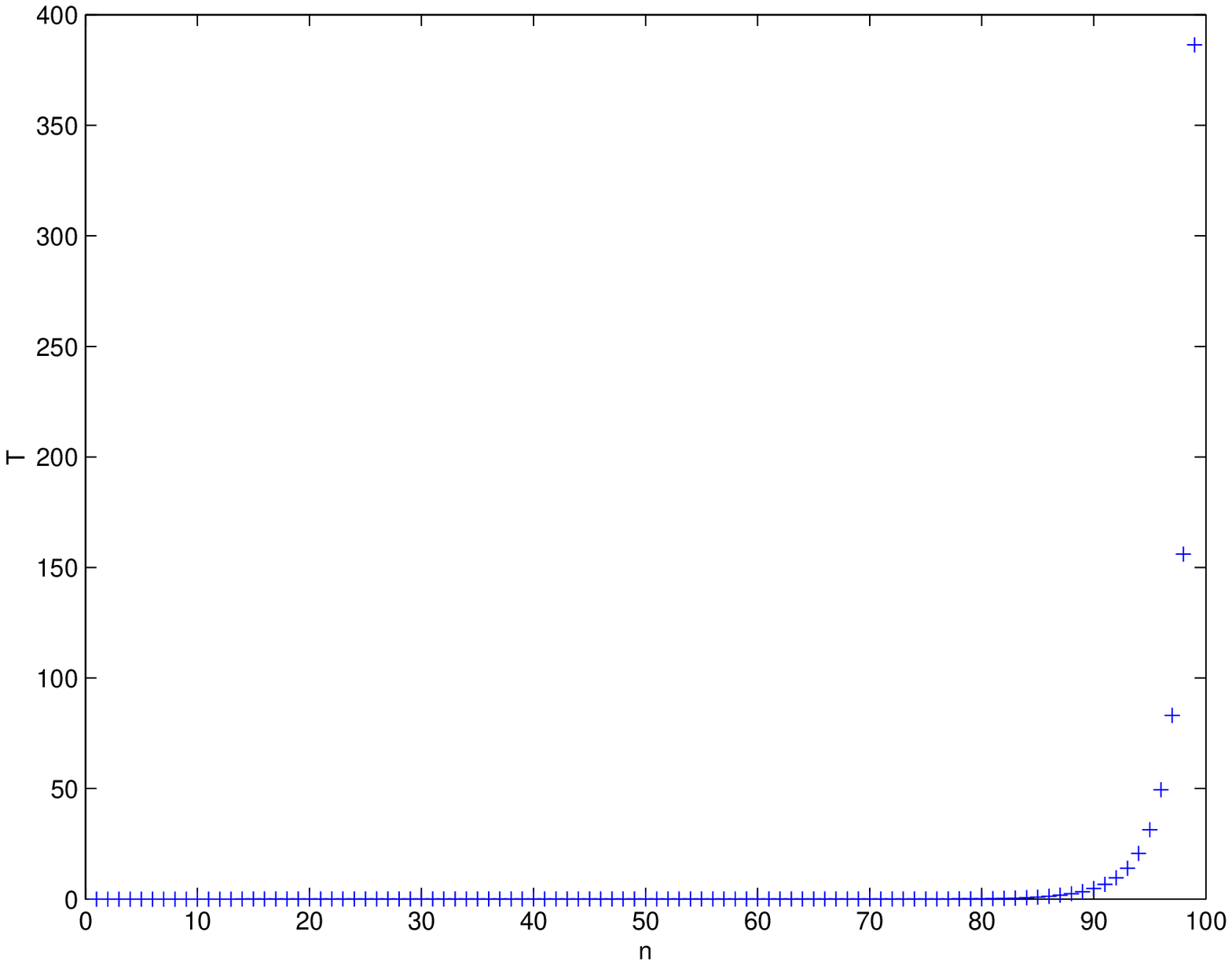, scale=0.4}{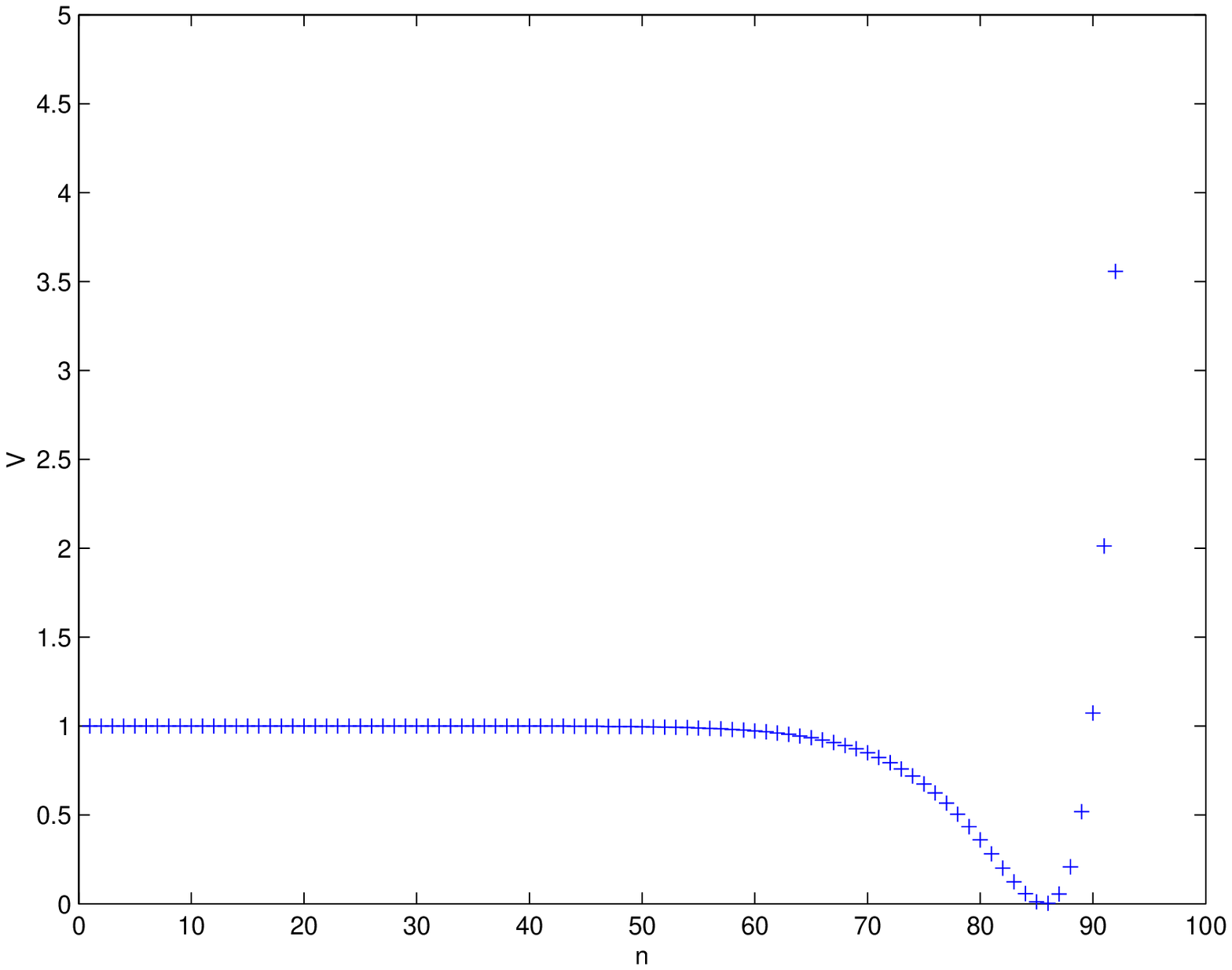, scale=0.4}{\label{K}Kinetic energy density for $M=N-1$ with $N=100$}{\label{P}Potential energy density for $M=N-1$ with $N=100$ (figure cut-off at $V< 5$)}  

The quasi-hole solutions that we have found are deformations of these ground states.  Comparing to the case of a quasi-hole with small $q$, we see from equations (\ref{C}) and (\ref{CT}) that the energy density perturbation is solely in the kinetic energy.  It is also localized at the origin as is evident from the form of the perturbation equation (\ref{P}).  The change in the energy is linear in $q$.  The solutions with large $q$ are not perturbative and should be rightly considered as solitons.  The question of the stability of our identified ground states against creation of solitons with large $q$ (the branch with decreasing enenrgy, see figure (\ref{energie cin fig})) is beyond the scope of this paper. 

The rotational excitations that we have found presumably give the Landau levels, upon quantization.  These are extremely widely separated in energy for large external magnetic field and hence will decouple from small excitations.  The ground state will correspond to the lowest Landau level.  The degeneracy of a given  Landau level is obtained from the total number of particles available and the physical space in which these particles are constrained.  The Landau levels will presumably play an important role in describing the transitions between different Hall plateaux.

\section{Conclusion}
In this article we have found a rich spectrum of the excitations of the
noncommutative droplet with an action based on the Chern-simons term and an
additional Maxwell term.  The existence of the Maxwell term seems to have
profound implications on the spectrum of fundamental oscillations.  We have
found two kind of solutions, the rotational
excitation and the quasi-hole solution.

The rotational excitations will exist in the finite or infinite case above
any solution of the equations of motion. The rotational frequency is exactly
the cyclotron frequency for the electrons in the model.   The rotational
excitation correspond to the energy of $N$ electrons moving in the magnetic
field when the amplitude of the rotational excitation is fixed to the radius
of the cyclotron motion.  With the quantization of these oscillations we
should recover the familiar Landau levels.

The quasi-holes solutions seem to exist only for the finite droplet.  The 
large
quasi-holes correspond to a thin annulus that undergoes nontrivial
oscillatory vibration.  They give a continuous band of solutions built above
each soliton solution of the field equations.  They also show a gap in their
spectrum as a function of the ``charge'' $q$ for small enough $\lambda$.

We expect that this spectrum of excitations will give rise to a complex
phenomenolgy which will allow us to describe transitions in two dimensional
magnetohydrodynamics, even the quantum Hall system.

\acknowledgments
We thank NSERC of Canada for financial support.

\end{document}